\documentstyle[12pt,subeqnarray,psfig]{article}
\begin{document}
%
%
\newenvironment{lefteqnarray}{\arraycolsep=0pt\begin{eqnarray}}
{\end{eqnarray}\protect\aftergroup\ignorespaces}
\newenvironment{lefteqnarray*}{\arraycolsep=0pt\begin{eqnarray*}}
{\end{eqnarray*}\protect\aftergroup\ignorespaces}
\newenvironment{leftsubeqnarray}{\arraycolsep=0pt\begin{subeqnarray}}
{\end{subeqnarray}\protect\aftergroup\ignorespaces}
\newcommand{\appleq}{\stackrel{<}{\sim}}
\newcommand{\appgeq}{\stackrel{>}{\sim}}
\newcommand{\arcsinh}{\mathop{\rm arcsinh}\nolimits}
\newcommand{\arctg}{\mathop{\rm arctg}\nolimits}
\newcommand{\diff}{{\rm\,d}}
\newcommand{\displayfrac}[2]{\frac{\displaystyle #1}{\displaystyle #2}}
\newcommand{\Erfc}{\mathop{\rm Erfc}\nolimits}
\newcommand{\Int}{\mathop{\rm Int}\nolimits}
\newcommand{\Nint}{\mathop{\rm Nint}\nolimits}
\newcommand{\pprime}{{\prime\prime}}

\title{Holes within galaxies: the  
       egg or the hen?}
\author{{R. Caimmi\footnote{
{\it Astronomy Department, Padua Univ., Vicolo Osservatorio 2,
I-35122 Padova, Italy} 
email: roberto.caimmi@unipd.it
}
\phantom{agga}}}

\maketitle
\begin{quotation}
\section*{}
\begin{Large}
\begin{center}

Abstract

\end{center}
\end{Large}
\begin{small}


Unsustained matter distributions unescapely collapse unless
fragmentation and centrifugal or pressure
support take place.   Starting from the above
evidence, supermassive compact objects at the
centre of large-mass galaxies (defined as
``holes'') are conceived as the end-product
of the gravitational collapse of local
density maxima (defined as ``central collapse'')
around which positive density perturbations
(overdensities) are located.   At the beginning
of evolution, assumed to occur at recombination
epoch, local density maxima are idealized as
homogeneous peaks, while the surrounding envelopes
are described by a power-law density profile,
$\rho(r)\propto r^{b-3}$, $0\le b\le3$, where
$b=0$ represents a massless atmosphere and
$b=3$ a homogeneous layer.   The dependence
of the density profile on a second parameter,
chosen to be the ratio between peak and total
(truncated) mass, $\kappa=M_{\rm pk}/M_{\rm tr}$,
is analysed.   Overdensity evolution is discussed
in the context of quintessence cosmological
models, which should be useful in dealing
with the virialized phase.   Aiming to describe
the central collapse, further investigation is
devoted to a special case where the quintessence
effect is equivalent to additional curvature
$(w=-1/3)$, and overdensities exhibit the
selected density profile at recombination epoch.
A redshift-dependent, power-law relation between
hole and (nonbaryonic) dark halo mass is
used to express the dependence of the fractional
mass, $\kappa$, on the overdensity mass,
$M=M_{\rm tr}$, where the homogeneous peak and
overdensity mass are related to the hole and
dark halo mass, respectively.   Computations
are performed for a wide range of masses,
$-1\le\log(M/{\rm M}_{10})\le6$, and mean overdensity
heights, $1\le\bar{\nu}_i\le4$, up to the
end of central collapse, and density profiles
of related configurations are determined
together with additional parameters.   The
central collapse is completed in early times,
no longer than a few hundredths of Gyr, which
implies hole formation when proto-haloes,
proto-bulges, and proto-disks are
still relaxing.   No appreciable
change in evolution (up to the end of central
collapse) is found with regard to
different mean overdensity heights related to equal
masses.   On the other hand, it is recognized
that homogeneous peaks collapse (in dimensionless
coordinates) ``faster'' with respect to
surroundings envelopes, in low-mass overdensities
than in large-mass overdensities.   In conclusion,
it is inferred that gravitational collapse of
homogeneous peaks within overdensities may be
a viable mechanism for hole generation.

\noindent
{\it keywords - Dark matter; Dark energy; 95.35.+d; 95.36+x.}

\end{small}
\end{quotation}

%

\section{Introduction}\label{intro}
The existence of large masses confined in a restricted
central region of galaxies, was first suggested in
order to explain short-period variability in brightness
exhibited by quasars and active galactic
nuclei (e.g., Terrel, 1964; Rees, 1966), and is widely
supported by current high-resolution observations
(for a review see e.g., Ferrarese and Ford, 2005;
Merritt, 2006).   There is an amount of increasing
evidence, that compact objects at the centre of
galaxies are
supermassive black holes (e.g., Maor, 1998; Miller,
2006), but their origin is still debated.

As a first alternative, supermassive black holes
are the end-product
of compact (high-density) stellar systems which make
galactic nuclei, where star encounters and star
collisions are the dominant physical processes
(Spitzer and Saslaw, 1966; Spitzer and Stone, 1967;
Colgate, 1967; Sanders, 1970).   Gravitational
scattering (via star-star encounters) of energetic
stars into elongated orbits, makes the density of
the central region increase which, in turn, implies
a higher rate of star evaporation, star collisions,
star disruption, and star coalescence.   The gas
produced during (high-energy) star collisions,
star disruption, and (high-mass) star death,
falls to the central region and condenses into
new stars which undergo further collisions.
Accordingly, the end-product of the evolution
of a dense nucleus appears to be the formation
of a central, massive black hole, either by
runaway star coalescence or gravitational
collapse of a massive gas cloud (Begalman
and Rees, 1978).   The possibility of seeds
for supermassive black holes which grow via
accretion of stars and/or gas liberated by
star collisions and star disruption, is also
considered (Duncan and Shapiro, 1983; David
et al., 1987a,b; Quinlan and Shapiro, 1987,
1989).   Seeds for supermassive black holes
could also be remnants of pop.\,III stars
(e.g., Madau and Rees, 2001; Johnson and
Bromm, 2007), or the (still speculative)
intermediate-mass black holes which form
in dense star cluster via merger events
(e.g., Portegies Zwart et al., 2004;
Maccarone et al., 2007).   For further
details, refer to Merritt (2006).

As a second alternative, supermassive black
holes take origin before or during galaxy
formation.   In the spinar model (Morrison,
1969; Cavaliere et al., 1969; Cavaliere et
al., 1970) the supermassive black hole
progenitor is a single object (spinar).
The spinar may be conceived as a
``gravitational machine'', where
gravitational energy is converted into
rotational energy via angular momentum
conservation during the collapse phase,
and rotational energy is converted into
nonthermal electromagnetic energy
(syncroton radiation) via acceleration
of charged particles (mainly electrons)
up to the relativistic regime, from
magnetic torques due to the surface
magnetic field.   According to recent
investigations, intermediate-mass black
holes acting as seeds for supermassive
black holes, could be formed from
low-angular momentum material in
primordial disks (Koushiappas et al.,
2004) or isothermal collapse of atomic
hydrogen gas $(M_{\rm H}=10^7$-$10^9
{\rm m}_\odot)$
within primordial dark matter haloes
(Spaan and Silk, 2006).

The energy released by the formation
and growth of supermassive black holes
must have had a major impact on how
gas cooled to form galaxies and galaxy
clusters (Silk and Rees, 1998).   In
particular, accretion onto supermassive
black holes provides the energy source
for active galactic nuclei which, in
turn, affects the evolution of galaxies
(Silk, 2005).   There is increasing
evidence that supermassive black holes
play an important role in the formation
and global evolution of galaxies and of
the intergalactic medium (Merloni et
al., 2005).

Seeds for pregalactic black holes could
be found much earlier, related to topological
defects during the inflation epoch, in
particular closed domain walls (Rubin
et al., 2001; Khlopov et al., 2002,
2005; for an extensive review, see
Khlopov and Rubin, 2004).   The resulting
mass spectrum extends over a wide range,
from superheavy to deeply subsolar black
holes, where the upper limit follows from
the condition that pieces of closed walls
do not dominate within horizon before the
whole wall enters it, while the lower
limit is given by the condition that the
black hole gravitational radius exceeds
the width of contracting domain wall.
In addition, low-mass objects are
concentrated around their most massive
counterparts within a cluster (e.g.,
Khlopov et al., 2005).

Seeds for supermassive black holes,
or supermassive black holes, can
form even in absence of baryons as
an inevitable consequence of relativistic
core collapse following the gravothermal
catastrophe, under the restrictive
assumption of self-interacting dark
matter haloes (Balberg and Shapiro, 2002).
The properties exhibited by self-interacting
dark matter are in some ways intermediate
between their counterparts related to hot
and cold dark matter, respectively.

The current attempt starts with the
working hypothesis, that supermassive
black holes are the first structures
formed within (sufficiently massive)
density perturbations (hereafter
quoted as ``overdensities'').
Similarly, about fifty years ago,
galaxies were conceived as emerging
from superdense primordial nuclei
(Ambartsumian, 1958, 1965).   In
fact, unsustained matter distributions
unescapely collapse unless fragmentation
and centrifugal or pressure support
take place.   To this respect, the
inner central region of collapsing
proto-galaxies is a very special
place, where both tidal effects and
acquisition of angular momentum are
minimized and may safely be thought
of as negligible.   Accordingly,
both spherical symmetry and homogeneity
are expected to be preserved,
which implies the formation
of a supermassive black hole, as in
large-mass $(m>20$\,-25m$_\odot)$
supernovae (e.g., Nomoto et al., 2005;
Ohkube et al., 2006).

If (nonbaryonic) dark matter is mainly
made of fermions (e.g., photinos), a
fermion ball instead of a black hole
could, in principle, take origin (e.g.,
Viollier, 1994; Munyaneza and Viollier,
2002; Munyaneza and Biermann, 2006)
but no firm conclusion can be reached
unless the nature of dark matter is
known.   Concerning ordinary matter,
gravitational collapse can be halted
by Fermi pressure for sufficiently
low masses (white dwarfs, neutron
stars, quark stars) but the contrary
holds for higher masses, where the
mean density of configurations close
to the gravitational i.e. Schwartzschild
radius is decreased.   If a similar
trend occurs for dark matter, and
supermassive black holes are the
result of dark matter collapse, a
lower mass limit may safely be
expected.

A short collapse time of inner,
densest regions within overdensities,
not exceeding a few hundredths of
Gyr, would be consistent with the
presence of a supermassive $(M=3\,
10^9\,{\rm m}_\odot)$ black
hole detected in the quasar SDSSJ
1148+5251, at a redshift $z=6.41$
corresponding to a cosmic age $t=
0.840$ Gyr (Willott et al., 2003).

Supermassive black hole formation
via gravitational collapse of the
local maxima around which overdensities
are placed (hereafter quoted as
``central collapse'') is a viable
scenario, to be tested when the
nature and the properties of
(nonbaryonic) dark matter are
known.   In this view, the current
paper investigates the special
configuration related to the end
of central collapse, for different
overdensity masses $(-1\le\log(M/
{\rm M}_{10})\le6$; ${\rm M}_{10}=10^{10}{\rm m}_
\odot)$ and mean peak heights
$(1\le\overline{\nu}_i\le4)$, with
assigned density profile at the
beginning of evolution.   Strictly
speaking, a change of the
quintessence equation of state
parameter, $w$, and the quintessence
degree of clustering, $\Gamma$,
should be taken
into consideration, in dealing
with overdensity evolution.
The general situation shall
widely be discussed, to show
how the above mentioned parameters
could affect the virialized
configuration.   
On the other hand, it will be
shown that central collapse is
completed in early times $(z>
10)$ which makes the effect of
quintessence negligible concerning
both the equation of state parameter
(e.g., Horellou and Berge, 2005) and
the degree of clustering (e.g., Nunes
and Mota, 2006).
Accordingly, detailed calculations
shall be limited to a special case
where considerable simplification
may be attained, overdensity
virialization being outside the
aim of the current attempt.

Some basic considerations are
presented in Sect.\,\ref{baco},
where the assumed initial
density profile is also defined.
The initial overdensity
configuration and related
evolution up to the end of
central collapse, are described
in Sect.\,\ref{evco}.   The
results are presented and
discussed in Sects.\,\ref{resu}
and \ref{disc}, respectively.
The conclusion makes the subject
of Sect.\,\ref{conc}.

\section{Basic considerations}
\label{baco}
\subsection{The hole}
\label{hole}

The existence of supermassive black holes
at the centre of large-mass star spheroids,
is supported with increasing evidence (e.g.,
Ferrarese and Ford, 2005; Merritt, 2006).
For sake of brevity, supermassive compact
objects within a restricted volume (in
particular, supermassive black holes)
shall be quoted as ``holes''.   Accordingly,
main components of a typical large-mass
galaxy are: the (nonbaryonic) dark halo,
the stellar halo, the disk, the bulge,
and the hole.   The hole, in turn, is
surrounded by an accretion disk, which
should be quoted as ``vortex'' or
``m{\ae}lmstrom'', and a dust torus,
with the emission of a jet along the
polar directions during an active
phase (e.g., Ferrarese and Ford, 2005).

\subsection{Black hole generation via
gravitational collapse}
\label{bhgc}

According to the density profile of
their progenitor, black hole formation
can occur bottom-up or top-down.   To
get more insight, let us refer to a
spherical-symmetric, truncated power-law
mass distribution, as:
\begin{leftsubeqnarray}
\slabel{eq:Mra}
&& \frac{M(r)}{M_{\rm{tr}}}=\left(\frac r{R_{\rm tr}}
\right)^b~~; \\
\slabel{eq:Mrb}
&& 0\le r\le R_{\rm tr}~~;\qquad M_{\rm tr}=M(R_{\rm tr})~~;
\qquad0\le b\le3~~;
\label{seq:Mr}
\end{leftsubeqnarray}
where $R_{\rm tr}$ is the truncation radius,
$M_{\rm tr}$ the mass within the truncation
radius, $b=0,1,3$ correspond to a Roche sphere
(central mass point surrounded by a massless
atmosphere), a (truncated) isothermal sphere,
and a homogeneous sphere, respectively.

Matter distributions expressed by Eq.\,(\ref
{seq:Mr}) are simple but nontrivial.   For
instance, models of baryonic collapse within
dark matter haloes and formation of gaseous
galactic disks involve use of truncated
power-law mass distributions (Adams and
Bloch, 2006).

The density profile related to Eq.\,(\ref
{seq:Mr}) reads:
\begin{leftsubeqnarray}
\slabel{eq:rora}
&& \frac{\rho(r)}{\rho_{\rm{tr}}}=\left(\frac r{R_{\rm tr}}
\right)^{b-3}~~;\qquad0\le r\le R_{\rm tr}~~; \\
\slabel{eq:rorb}
&& \rho_{\rm tr}=\rho(R_{\rm tr})=\frac b3\overline{\rho}_
{\rm tr}~~;\qquad\overline{\rho}_{\rm tr}=\frac3{4\pi}\frac
{M_{\rm tr}}{R_{\rm tr}^3}~~;
\label{seq:ror}
\end{leftsubeqnarray}
where $\rho_{\rm{tr}}$ and $\overline
{\rho}_{\rm tr}$ are the local density
at the truncation radius and the mean
density within the truncation radius,
respectively.   The central density
cusp can be avoided by replacing an
arbitrarily small inner core at $r=r_
{\rm pk}$, with a homogeneous core where
$\rho(r)=\rho(r_{\rm pk}),$ $0\le r\le
r_{\rm pk}.$

A spherical-symmetric matter distribution
collapses into a black hole when the
gravitational i.e. Schwartzschild radius
(e.g., Landau and Lifchitz, 1966, Chap.\,XI,
\S\,97) is attained, as:
\begin{equation}
\label{eq:ragv}
R_{\rm gr}=\frac{2GM_{\rm tr}}{c^2}~~;
\end{equation}
where $G$ is the constant of gravitation
and $c$ the light velocity in vacuum, to be
intended as the light velocity in absence of
baryonic matter.   For further details,
refer to Appendix \ref{a:c}.

For the density profiles under discussion,
the gravitational radius related to a sphere
of radius, $r$, enclosing a mass, $M(r)$, is
obtained by the combination of Eqs.\,(\ref
{seq:Mr}) and (\ref{eq:ragv}), the latter
generalized to a mass, $M(r)$.   The result is:
\begin{equation}
\label{eq:rgRg}
\frac{r_{\rm gr}}{R_{\rm gr}}=\left(\frac rR\right)^b~~;
\end{equation}
where $R_{\rm gr}$ is the gravitational radius
related to the truncation mass, according to
Eq.\,(\ref{eq:ragv}).

In the special case of the isothermal sphere,
$b=1$, all the spheres collapse into a black
hole at the same time.   Milder density
profiles, $1<b\le3$, imply $r_{\rm gr}/r<
R_{\rm gr}/R_{\rm tr}$ i.e. top-down black
hole formation: the gravitational radius is
first attained at the truncation radius.
Harder density profiles, $0\le b<1$, imply
$r_{\rm gr}/r>R_{\rm gr}/R_{\rm tr}$ i.e.
bottom-up black hole formation: the
gravitational radius is first attained at
the centre.

If holes are the end-products of overdensity
central collapse, the formation mechanism
must necessarily be bottom-up with regard
to the whole mass, but all the possibilities
remain open with regard to the local maximum,
conceived as the proto-hole.

\subsection{Initial density profiles}
\label{idep}

Power-law density profiles provide a poor
fit to dark matter haloes unless sufficiently
thin shells are considered.   A better
description relies on three-exponent laws
with a central cusp (e.g., Hernquist, 1990;
Navarro et al., 1995, 1996; Zhao, 1996; Moore
et al., 1998, 1999; Fukushige and Makino,
2001, 2003; Diemand et al., 2004; Reed et
al., 2005; for the description of a fitting
procedure see e.g., Caimmi and Marmo, 2003,
2004; Caimmi et al., 2005; Caimmi, 2006) or
two-parameter
laws (e.g., Sersic, 1968; Navarro et al.,
2004; Merritt et al., 2005).

On the other hand, it is not the case at the
beginning of the evolution which, for high-energy
universes, may safely be related to recombination
epoch (e.g., Caimmi, 1989), where the contribution
of dark energy is still negligible.    Aiming to
describe overdensity evolution up to the end of
the central collapse, a simple but nontrivial
density profile is assumed.

More specifically, overdensities at recombination
epoch $(z\approx1100)$ are represented as
homogeneous peaks surrounded by envelopes
where the density profile is represented
by a power-law, according to Eqs.\,(\ref
{seq:ror}).  The explicit expression reads:
\begin{equation}
\label{eq:ropi}
\rho(r)=\cases{
\rho_{\rm pk}~~; & $0\le r\le r_{\rm pk}~~;$ \cr
\rho_{\rm pk}\left(\frac r{r_{\rm pk}}\right)^{b-3}~~; &
$r_{\rm pk}\le r\le R_{\rm tr}~~;$ \cr}
\end{equation}
where $\rho_{\rm pk}$ and $r_{\rm pk}$ are
the homogeneous peak density and radius,
respectively.

Owing to spherical symmetry, the mass
enclosed within the truncation surface is:
\begin{equation}
\label{eq:MR}
M_{\rm tr}=\int_0^{R_{\rm tr}}\rho(r)4\pi r^2\diff r~~;
\end{equation}
which may be integrated using Eq.\,(\ref
{eq:ropi}).   The result is:
\begin{leftsubeqnarray}
\slabel{eq:MRCa}
&& M_{\rm tr}=M_{\rm pk}\left\{1+\frac3b\left[
\left(\frac{R_{\rm tr}}{r_{\rm pk}}\right)^b-1
\right]\right\}~~; \\
\slabel{eq:MRCb}
&& M_{\rm pk}=\frac{4\pi}3\rho_{\rm pk}r_{\rm pk}^3~~;
\label{seq:MRC}
\end{leftsubeqnarray}
where $M_{\rm pk}$ is the homogeneous peak
mass.

Using Eq.\,(\ref{eq:MRCa}), a dimensionless
radius, $\Xi$, may be defined in terms of a
dimensionless mass, $\kappa$, as:
\begin{leftsubeqnarray}
\slabel{eq:Rrca}
&& \Xi=\left(1+\frac b3\frac{1-\kappa}\kappa\right)^
{1/b}~~; \\
\slabel{eq:Rrcb}
&& \Xi=\frac{R_{\rm tr}}{r_{\rm pk}}~~;\qquad
\kappa=\frac{M_{\rm pk}}{M_{\rm tr}}~~;
\label{seq:Rrc}
\end{leftsubeqnarray}
and the mean density within the truncation
radius, $\overline{\rho}_{\rm tr}=(3/4\pi)
(M_{\rm tr}/R_{\rm tr}^3)$, owing to Eqs.\,(\ref
{seq:MRC}) and (\ref{seq:Rrc}), takes the
expression:
\begin{equation}
\label{eq:romR}
\frac{\overline{\rho}_{\rm tr}}{\rho_{\rm pk}}=\Xi^{-3}
\left[1+\frac3b\left(\Xi^b-1\right)\right]~~;
\end{equation}
in terms of the dimensionless radius, $\Xi$, and:
\begin{equation}
\label{eq:romc}
\frac{\overline{\rho}_{\rm tr}}{\rho_{\rm pk}}=\frac1\kappa
\left[1+\frac b3\frac{1-\kappa}\kappa\right]^{-3/b}~~;
\end{equation}
in terms of the dimensionless mass, $\kappa$.

The combination of Eqs.\,(\ref{eq:ropi}) and (\ref{seq:Rrc})
yields the local density at the truncation radius, as:
\begin{equation}
\label{eq:roRc}
\frac{\rho_{\rm tr}}{\rho_{\rm pk}}=
\left[1+\frac b3\frac{1-\kappa}\kappa\right]^{(b-3)/b}~~;
\end{equation}
and the combination of Eqs.\,(\ref{eq:romc}) and
(\ref{eq:roRc}) yields the local to global density
ratio at the truncation radius, as:
\begin{equation}
\label{eq:rorR}
\frac{\rho_{\rm tr}}{\overline{\rho}_{\rm tr}}=
\kappa\left[1+\frac b3\frac{1-\kappa}\kappa\right]~~;
\end{equation}
which depend on two parameters, $b$ and $\kappa$.

With regard to a generic dimensionless radius,
$\xi$, $\xi_{\rm pk}\le\xi\le\Xi$, Eqs.\,(\ref
{seq:MRC}), (\ref{eq:romR}), and (\ref{eq:ropi})
may be generalized as:
\begin{lefteqnarray}
\label{eq:Mr}
&& M(r)=M_{\rm pk}\left[1+\frac3b\left(\xi^b-1\right)\right]~~; \\
\label{eq:romr}
&& \overline{\rho}(r)=\frac{\rho_{\rm pk}}{\xi^3}
\left[1+\frac3b\left(\xi^b-1\right)\right]~~; \\
\label{eq:ror}
&& \rho(r)=\rho_{\rm pk}\xi^{b-3}~~;\qquad\xi=\frac r{r_{\rm pk}}~~;
\end{lefteqnarray}
and the combination of Eqs.\,(\ref{eq:romr}) and
(\ref{eq:ror}) yields:
\begin{equation}
\label{eq:rorm}
\frac{\rho(r)}{\overline{\rho}(r)}=
\left[\frac3b+\frac{b-3}b\xi^{-b}\right]^{-1}~~;
\end{equation}
which is the generalization of Eq.\,(\ref
{eq:rorR}) to a generic dimensionless radius,
$\xi$, $\xi_{\rm pk}\le\xi\le\Xi$.

For a complete definition of density
profiles within overdensities at recombination
epoch, the boundary conditions derived from
the cosmological model are needed.

\section{Overdensity evolution up to the end
of central collapse}
\label{evco}

Let overdensities be made of a homogeneous
peak surrounded by a heterogeneous envelope,
at recombination epoch.   The validity of the
above mentioned assumption implies hole
formation unless the collapse is halted
by Fermi pressure, as in neutron star
progenitors.   In absence of any information
about the (nonbaryonic) dark matter, hole
formation via gravitational collapse cannot
be escluded and must be taken into consideration.
To this aim, overdensity evolution up to the end
of central collapse shall be investigated in
what follows.

Given a spherical-symmetric overdensity around
a local maximum at the beginning of evolution,
the central collapse is intended as the collapse
of a dusty homogeneous sphere with same density
and expansion rate with respect to the local
maximum.

\subsection{Overdensities at recombination epoch}
\label{defr}

The effect of dark energy may safely be
neglected in early times $(z>10)$ and
matter may be described, to a good extent
after recombination,
in the context of Friedmann universes with
dust only.   With regard to overdensities
bound around a local density maximum, or
peak (e.g., Heavens and Peacock, 1998),
related mean and local values are
defined as:
\begin{equation}
\label{eq:delRr}
\overline{\delta}=\frac{\overline{\rho}}{\rho_{hm}}-1~~;\qquad
\delta=\frac\rho{\rho_{hm}}-1~~;
\end{equation}
where $\rho_{hm}$ is the mean density of the
matter Hubble flow.   Keeping in mind the
definition of mean density, the combination
of Eqs.\,(\ref{eq:delRr}) yields (e.g.,
Ryden and Gunn, 1987):
\begin{equation}
\label{eq:delin}
\overline{\delta}=\frac1S\int\int\int_S\delta\diff^3S~~;
\end{equation}
where $S$ is the overdensity volume.

For high-energy universes, as in current
QCDM cosmologies where $\Omega_m\approx
0.3$, $\Omega_q\approx0.7$, $\Omega_m+
\Omega_q=1$, the initial configuration
may safely be related to the recombination
epoch (see e.g., Caimmi, 1989 for the
special case of CDM cosmologies).
According to Eq.\,(\ref{eq:delRr}),
the overdensity mass reads:
\begin{equation}
\label{eq:M}
M=(1+\overline{\delta}_i)\rho_{hm}(a_{hi})S_i=
(1+\overline{\delta}_i)M_{hm}~~;
\end{equation}
where $M_{hm}$ is the mass of matter Hubble
flow enclosed within the volume, $S_i$,
and the index, $i$, denotes recombination.

The mean overdensity value can be expressed
in terms of the peak overdensity value,
$\delta_{\rm pk}$, as:
\begin{equation}
\label{eq:delm}
\overline{\delta}=F\delta_{\rm pk}~~;\qquad
F=\frac1S\int\int\int_S\frac{\delta(\vec{r})}
{\delta_{\rm pk}}\diff^3S~~;
\end{equation}
where the shape factor, $F$, lies in the
range, $S_{\rm pk}/S\le F\le1$, between
the extreme situations of a peak
surrounded by a massless atmosphere
(lower limit) and by a layer of
equal density (upper limit); and $S_
{\rm pk}$ represents the volume filled
by the peak, assumed to be homogeneous.

Let the mean and the local overdensity
height be defined as the ratios:
\begin{equation}
\label{eq:numl}
\overline{\nu}=\frac{\overline{\delta}}
{\overline{\delta}_{\rm M}}~~;\qquad\nu=\frac\delta
{\overline{\delta}_{\rm M}}~~;\qquad{\overline{\delta}_{\rm M}}
=<\overline{\delta}^2>_{\rm M}^{1/2}~~;
\end{equation}
where $<\overline{\delta}^2>_{\rm M}^{1/2}$ is the
rms overdensity value related to an assigned
volume, $S$, at a given epoch (recombination
in the case under discussion), and the index,
$M$, means that a volume, $S$, within the
Hubble flow, encloses a mass, $M=M_{hm}=\rho_{hm}S$.
The combination of Eqs.\,(\ref{eq:delm})
and (\ref{eq:numl}) yields:
\begin{equation}
\label{eq:num}
\overline{\nu}=F\nu_{\rm pk}~~;\qquad\nu_{\rm pk}=
\frac{\delta_{\rm pk}}{\overline{\delta}_{\rm M}}~~.
\end{equation}

If, in addition, the density profile is
expressed by Eq.\,(\ref{eq:ropi}), the
combination of Eqs.\,(\ref{eq:romr}),
(\ref{eq:ror}), (\ref{eq:rorm}), and
(\ref{eq:delRr}) yields:
\begin{lefteqnarray}
\label{eq:derm}
&& 1+\overline{\delta}(r)=\frac{1+\delta_{\rm pk}}{\xi^3}
\left[1+\frac3b\left(\xi^b-1\right)\right]~~; \\
\label{eq:der}
&& 1+\delta(r)=(1+\delta_{\rm pk})\xi^{b-3}~~; \\
\label{eq:debrm}
&& 1+\delta(r)=[1+\overline{\delta}(r)]
\left[\frac3b+\frac{b-3}b\xi^{-b}\right]^{-1}~~; \\
\label{eq:dec}
&& \delta_{\rm pk}=\frac{\rho_{\rm pk}-\rho_{hm}}{\rho_{hm}}~~;
\qquad1\le\xi\le\Xi~~;
\end{lefteqnarray}
and the particularization to the
truncation radius, $\xi=\Xi$, by
use of Eqs.\,(\ref{eq:MRCa}) and
(\ref{seq:Rrc}) produces:
\begin{lefteqnarray}
\label{eq:deRm}
&& 1+\overline{\delta}_{\rm tr}=\frac{1+\delta_{\rm pk}}\kappa
\left[1+\frac b3\frac{1-\kappa}\kappa\right]^{-3/b}~~; \\
\label{eq:deR}
&& 1+\delta_{\rm tr}=(1+\delta_{\rm pk})\left[1+\frac b3
\frac{1-\kappa}\kappa\right]^{1-3/b}~~; \\
\label{eq:debRm}
&& 1+\delta_{\rm tr}=(1+\overline{\delta}_{\rm tr})\kappa
\left[1+\frac b3\frac{1-\kappa}\kappa\right]~~; \\
\label{eq:lib0}
&& \lim_{b\to0}\left[1+\frac b3\frac{1-\kappa}\kappa\right]^
{1-3/b}=\exp\left(-\frac{1-\kappa}\kappa\right)~~; \\
\label{eq:dib0}
&& 0<\left[1+\frac b3 \frac{1-\kappa}\kappa\right]^
{1-3/b}\le1~~;\qquad0<b\le3~~;
\end{lefteqnarray}
accordingly, overdensities at recombination are 
defined by the following parameters: total mass,
$M=M_{\rm tr}$; mean value, $\overline{\delta}=
\overline{\delta}_{\rm tr}$; mean height,
$\overline{\nu}=\overline{\nu}_{\rm tr}$;
fractional mass, $\kappa$; power-law density
profile exponent, $b$; which allow the calculation
of the peak overdensity, $\delta_{\rm pk}$,
and the local overdensity, $\delta(r)$.   At
this stage, a cosmological model needs to be
specified.

\subsection{Input parameters and cosmological model}
\label{ipcm}

Overdensity masses and mean heights shall be
considered in the range, $-1\le\log(M/{\rm M}_{10})
\le6$ $({\rm M}_{10}=10^{10}{\rm m}_\odot)$ and $1
\le\overline{\nu}\le4$, respectively.   Present-day rms
overdensity values in the linear approximation,
shall be assumed as in CDM cosmologies (e.g.,
Gunn, 1987; Ryden and Gunn, 1987).   Related
values at recombination may be determined,
provided the temporal behaviour of the rms
overdensity value for a selected mass, is
known.

For QCDM cosmologies, it has still to be
found an analytical solution which represents
the growing mode of overdensities.   In the
special case of flat universes, $\Omega_m+
\Omega_q=1$, and time-independent quintessence
equation of state parameter, $w={\rm const}$,
the solution may be written in terms of the
hypergeometric function, $_2F_1$ (Silveria
and Wega, 1994).

Extending the results related to the special
case where the dark energy mimics a cosmological
constant, $w=-1$ (Wang and Steinhardt, 1998;
Carroll et al., 1992) to the range, $w>-1$
(Basilakos, 2003), and improving the results
related to phantom energy, $w<-1$, yield the
following approximation (Percival, 2005):
\begin{leftsubeqnarray}
\slabel{eq:deva}
&& \frac{\overline{\delta}_{\rm M}(a_h)}{\overline
{\delta}_{\rm M}(1)}=\frac52\frac{\Omega_m
(a_h)a_h}{\left[\Omega_m(a_h)\right]^u-\Omega_q
(a_h)+\left[1+0.5\Omega_m(a_h)\right]\left[1+
\lambda\Omega_q(a_h)\right]}~~; \\
\slabel{eq:devb}
&& \Omega_m(a_h)=\frac{\Omega_ma_h^{-3}}
{\Omega_ma_h^{-3}+\Omega_qa_h^{f(a_h)}}~~; \\
\slabel{eq:devc}
&& \Omega_q(a_h)=\frac{\Omega_qa_h^{f(a_h)}}
{\Omega_ma_h^{-3}+\Omega_qa_h^{f(a_h)}}~~; \\
\slabel{eq:devd}
&& f(a_h)=-\frac3{\ln(a_h)}\int_0^{\ln(a_h)}
[1+w(x)]\diff\ln(x)~~; \\
\slabel{eq:deve}
&& u=\frac{3(1-w)}{5-6w}\left\{1+\frac12\frac
{2-3w}{(5-6w)^2}[1-\Omega_m(a_h)]\right\}~~; \\
\slabel{eq:devf}
&& \lambda=-\frac1{10}\frac{75w+76}{25w+2}~~; \\
\slabel{eq:devg}
&& \Omega_m(a_h)+\Omega_q(a_h)=1~~;\qquad
\Omega_m=\Omega_m(1)~~;\qquad\Omega_q=
\Omega_q(1)~~; \\
\slabel{eq:devh}
&& w(a_h)={\rm const}=w~~;
\label{seq:dev}
\end{leftsubeqnarray}
where $a_h$ is the cosmological scale factor,
normalized to $a_h=1$ at the present time,
$\Omega_m$ and $\Omega_q$ are density parameters
related to matter (including radiation in the
case under discussion, $z\appleq1100$) and
quintessence, respectively.    In the limit
of constant quintessence equation of state
parameter, Eq.\,(\ref{eq:devd}) reduces to
(e.g., Horellou and Berge, 2005):
\begin{equation}
\label{eq:fah}
f(a_h)=-3(1+w)~~;
\end{equation}
according to Eq.\,(\ref{eq:devh}).

At recombination epoch, $z=1100$ to a
good extent, and the cosmological scale
factor reads:
\begin{equation}
\label{eq:ahi}
(a_h)_i=\frac1{1+z_i}=\frac1{1101}~~;
\end{equation}
where the index, $i$, denotes the recombination.

The input parameters of the cosmological model
are assumed to be the following (e.g., Manera
and Mota, 2006):
\begin{equation}
\label{eq:copa}
\Omega_m=0.3~~;\qquad\Omega_q=0.7~~;\qquad
\Omega_b=0.047~~;\qquad h=0.65~~;\qquad\sigma_8=0.9~~;
\end{equation}
where $\Omega_b$ is the baryon density
parameter, $h=H/(100\,{\rm km\,s}^{-1}
{\rm Mpc}^{-1})$ is the dimensionless
Hubble parameter at $z=0$, and
$\sigma_8=\overline{\delta}_{M(8)}$
is related to the mass, $M(8)$, within
a sphere of radius, $R(8)=8h^{-1}$\,Mpc,
filled by Hubble flow at present.

A different dependence, $\sigma_8=f(w)$,
is deduced according if X-ray cluster
data (Wang and Steinhardt, 1998) or
cosmic microwave background data
(Doran et al., 2001) are used.   The
cases, $\sigma_8=0.8$ and $\sigma_8=0.7$
should also be considered (e.g., Horellou
and Berge, 2005), but a complete investigation
on this point is outside the aim of the
current paper.

The system of measure, [kpc M$_{10}$ Gyr],
M$_{10}=10^{10}$m$_\odot$, shall be adopted 
in performing computations.   The conversion
formulae from and towards the standard
astrophysical system of measure, [cm g s],
are reported in Appendix \ref{a:syme}.

Using Eqs.\,(\ref{eq:copa}), the current value
of critical and mean matter density are:
\begin{leftsubeqnarray}
\slabel{eq:roca}
&& \rho_{\rm crit}=\frac{3H^2}{8\pi G}=
1.172\,6\,10^{-8}{\rm M}_{10}{\rm kpc}^{-3}~~; \\
\slabel{eq:rocb}
&& \rho_{hm}=\Omega_m\rho_{\rm crit}=
3.517\,8\,10^{-9}{\rm M}_{10}{\rm kpc}^{-3}~~;
\label{seq:roc}
\end{leftsubeqnarray}
and the amount of matter Hubble flow
enclosed within a sphere of radius,
$R(8)=8h^{-1}$Mpc, is:
\begin{equation}
\label{eq:M8}
M(8)=\frac{4\pi}3\rho_{hm}\left(\frac8h\right)^3=
94402{\rm M}_{10}~~;
\end{equation}
which is consistent with a value $M(8)=
932513{\rm M}_{10}$ deduced by use of a
CDM cosmological model (Gunn, 1987).
The rms overdensity spectrum at recombination
epoch, $\overline{\delta}_{\rm M}(a_{hi})$,
shall be deduced from the above mentioned
CDM model (e.g., Caimmi et al., 1990).
Related present-day values in the linear
theory, $\overline{\delta}_{\rm M}(1)$,
are deduced from the standard relation
(e.g., Zeldovich and Novikov, 1982):
\begin{equation}
\label{eq:rhoah}
\rho_{hm}(a_h)=\rho_{hm}(1)(1+z)^3~~;
\end{equation}
where $z$ is the redshift.

The radius of a sphere filled by matter
Hubble flow of mass, $M=10^\ell{\rm M}_
{10}$, is defined by the relation:
\begin{equation}
\label{eq:Rhm}
M=10^\ell{\rm M}_{10}=\rho_{hm}(a_h)\frac{4\pi}3R^3(a_h)~~;
\end{equation}
and the combination of Eqs.\,(\ref{seq:roc})
and (\ref{eq:Rhm}) yields:
\begin{equation}
\label{eq:R0}
R_0=R(1)=0.189\,327\,10^{(\ell+10)/3}{\rm kpc}~~;
\end{equation}
with regard to the current epoch,
where $-1\le\ell\le6$ in the range
of interest.

A linear dependence, $R\propto a_h$,
implies the following:
\begin{equation}
\label{eq:Rah}
R(a_h)=\frac{R_0}{1+z}~~;
\end{equation}
at any selected time; in particular,
$R_i=R_0/1101$ at recombination epoch.

For nonrotating, spherical-symmetric
overdensities with null peculiar
velocity field, the turnaround radius,
$r_{\rm max}$, is defined by the relation
(e.g., Peebles, 1980, Chap.\,II, \S\,19A;
Caimmi, 1989):
\begin{leftsubeqnarray}
\slabel{eq:Rmaxa}
&& \frac{R_i}{r_{\rm max}}=\Delta_i~~; \\
\slabel{eq:Rmaxb}
&& \Delta_i=1-(\Omega_m)_i^{-1}(1+\overline{\delta}_i)^{-1}>0~~;
\label{seq:Rmax}
\end{leftsubeqnarray}
which holds for CDM universes.
The combination of Eqs.\,(\ref
{eq:numl}) and (\ref{seq:Rmax}) yields:
\begin{equation}
\label{eq:Rnmx}
\overline{\nu}_ir_{\rm max}=R_i\left(\overline{\delta}_
{{\rm M}i}\right)^{-1}~~;\qquad\left\vert1-(\Omega_m)_i^
{-1}\right\vert\ll\overline{\delta}_i\ll1~~;
\end{equation}
to the first order in $\overline{\delta}_i$.

The rms overdensity value in the linear
theory, $\overline{\delta}_{{\rm M}i}$,
and the radius of the related sphere,
$R$, are listed in Tab.\,{\ref{t:CDM}
with regard to current and recombination
epoch, using a CDM model (Gunn, 1987),
as a function of the mass.
\begin{table}
\begin{tabular}{llllll}
\hline
\hline
\multicolumn{1}{c}{$\log(M/{\rm M}_{10})$} &
\multicolumn{1}{c}{$\overline{\delta}_{{\rm M}0}$} &
\multicolumn{1}{c}{$R_0$/kpc}
&
\multicolumn{1}{c}{$\overline{\delta}_{{\rm M}i}$} &
\multicolumn{1}{c}{$R_i$/kpc} &
\multicolumn{1}{c}{$\overline{\nu}_ir_{\rm max}$} \\
\hline
$-$1 & 14 & 1.89 E+2 & 1.27 E$-$2 
& 1.72 E$-$1 & 8.13 E+0 \\
$\phantom{-}0$ & 11 & 4.08 E+2 & 1.00 E$-$2 
& 3.70 E$-$1 & 2.23 E+1 \\
$\phantom{-}1$ & \phantom{1}8.0 & 8.79 E+2 & 7.26 E$-$3 
& 7.98 E$-$1 & 6.60 E+1 \\
$\phantom{-}2$ & \phantom{1}5.3 & 1.89 E+3 & 4.81 E$-$3 
& 1.72 E+0 & 2.15 E+2 \\
$\phantom{-}3$ & \phantom{1}3.4 & 4.08 E+3 & 3.08 E$-$3 
& 3.70 E+0 & 7.21 E+2 \\
$\phantom{-}4$ & \phantom{1}1.9 & 8.79 E+3 & 1.72 E$-$3 
& 7.98 E+0 & 2.78 E+3 \\
$\phantom{-}5$ & \phantom{1}0.87 & 1.89 E+4 & 7.89 E$-$4 
& 1.72 E+1 & 1.31 E+4 \\
$\phantom{-}6$ & \phantom{1}0.32 & 4.08 E+4 & 2.90 E$-$4 
& 3.70 E+1 & 7.66 E+4 \\
\hline\hline
\end{tabular}
\caption{The rms overdensity value in the linear
theory, $\overline{\delta}_{\rm M}$,
and the radius of the related sphere,
$R$, with regard to current and recombination
epoch, using a CDM model (Gunn, 1987),
as a function of the mass.   The product,
$\overline{\nu}_ir_{\rm max}$, deduced from
Eq.\,(\ref{eq:Rnmx}), is also listed.   The
current and recombination epoch are denoted
by the indices, 0 and $i$, respectively.
Overdensity values at recombination epoch,
$\overline{\delta}_{{\rm M}i}$, deduced from
their present-day counterparts, $\overline
{\delta}_{{\rm M}0}$, by use of Eqs.\,(\ref
{seq:dev}), related to QCDM models with
constant quintessence equation of state
parameter, $w$, yield $z_i=1101.526\,3$ instead
of the assumed value, $z_i=1100$, making a
close agreement, as expected.}
\label{t:CDM}
\end{table}
The product, $\overline{\nu}_ir_{\rm max}$,
deduced from Eq.\,(\ref{eq:Rnmx}), is also
listed therein.   
Overdensity values at recombination epoch,
$\overline{\delta}_{{\rm M}i}$, deduced from
their present-day counterparts, $\overline
{\delta}_{{\rm M}0}$, by use of Eqs.\,(\ref
{seq:dev}), related to QCDM models with
constant quintessence equation of state
parameter, $w$, yield $z_i=1101.5263$ instead
of the assumed value, $z_i=1100$, making a
close agreement, as expected.

Turning to the general case of QCDM cosmologies,
the dynamical expansion of the universe is
described by the Friedmann equations (e.g.,
Percival, 2005):
\begin{leftsubeqnarray}
\slabel{eq:Frieda}
&& \frac{H^2(a_h)}{H_0^2}=\Omega_ma_h^{-3}+\Omega_ka_h^{-2}+
\Omega_qa_h^{f(a_h)}~~; \\
\slabel{eq:Friedb}
&& \Omega_k=1-\Omega_m-\Omega_q~~;\qquad\Omega_u(a_h)=\frac
{\rho_u(a_h)}{\rho_{\rm crit}(a_h)}~~;\qquad u=m,q~~; \\
\slabel{eq:Friedc}
&& \rho_{\rm crit}(a_h)=\frac{3H^2(a_h)}{8\pi G}~~;\qquad
H(a_h)=\frac{\dot{a}_h}{a_h}~~;
\label{seq:Fried}
\end{leftsubeqnarray}
where $\Omega_k$ is the curvature density parameter,
which is null for flat universes, and Eqs.\,(\ref
{seq:dev})} hold.   The function, $f(a_h)$, is
calculated by solving the conservation of energy
equation for the dark energy (Caldwell et al., 1998a):
\begin{equation}
\label{eq:fahp}
\frac{\diff(c^2\rho_qa_h^3)}{\diff a_h}=-3p_qa_h^2~~;
\end{equation}
which yields:
\begin{equation}
\label{eq:rhoq}
\rho_q(a_h)=\rho_qa_h^{f(a_h)}~~;
\end{equation}
on the other hand, the conservation of matter mass
produces:
\begin{equation}
\label{eq:rhom}
\rho_m(a_h)=\rho_ma_h^{-3}~~;
\end{equation}
in the case under consideration of flat universes,
the combination of Eqs.\,(\ref{seq:Fried}),
(\ref{eq:rhoq}), and (\ref{eq:rhom}) yields:
\begin{equation}
\label{eq:Hrmq}
\left(\frac{\dot{a}_h}{a_h}\right)^2=H^2(a_h)=\frac{8\pi G}3
\left[\rho_ma_h^{-3}+\rho_qa_h^{f(a_h)}\right]~~;
\end{equation}
where the function, $f(a_h)$, is defined by
Eq.\,(\ref{eq:devd}), or Eq.\,(\ref{eq:fah})
in the limit of constant quintessence equation
of state parameter, $w$.

Though the action of the dark energy may be
explained in different ways (e.g., Caldwell
et al., 1998a), the current interpretation is
based on a dynamical evolving scalar field
slowly rolling down its potential, ${\cal V}
(\phi)$.   If quintessence is minimally
coupled to gravity i.e. any couplings to
other fields are supposed to be negligibly
small, the equation of motion for the scalar
field within the Hubble flow is (e.g., Manera
and Mota, 2006):
\begin{equation}
\label{eq:sfme}
\ddot{\phi}=-3H\dot{\phi}-\frac{\diff{\cal V}}{\diff\phi}~~;
\end{equation}
where the variable, $\phi$, may be conceived
as a curvilinear coordinate (with the dimension
of a length) which describes the evolution of
the scalar field.   A dynamical analogon of
Eq.\,(\ref{eq:sfme}) is e.g., a solid shpere
in free fall within a homogeneous fluid under
the action of gravitation, where
the first and the second term on the right-hand
side of Eq.\,(\ref{eq:sfme}) are related to
viscous and gravitational force, respectively.

The pressure and the density of the scalar
field are (e.g., Nunes and Mota, 2006):
\begin{lefteqnarray}
\label{eq:pq}
&& p_q=\frac12\rho_{\rm K}\dot{\phi}^2-\rho_{\rm P}{\cal V}(\phi)~~; \\
\label{eq:roq}
&& \rho_q=\frac12\frac{\rho_{\rm K}}{c^2}\dot{\phi}^2+\frac{\rho_{\rm P}}
{c^2}{\cal V}(\phi)~~;
\end{lefteqnarray}
where $\rho_{\rm K}$ and $\rho_{\rm P}$ are mass densities
related to kinetic and potential energy,
respectively, which implies $\rho_{\rm K}\dot{\phi}^2/2$
and $\rho_{\rm P}{\cal V}(\phi)$ are kinetic and
potential energy densities, respectively,
of the scalar field%
\footnote{The mass densities, $\rho_{\rm K}$ and
$\rho_{\rm P}$,
are usually omitted in literature under the pretext
that the chosen units imply unitary values of the
light velocity in vacuum, $c=1\,u_c$, and the Planck
constant, $h=1\,u_h$, where $u_c$ and $u_h$ are the
unit velocity and the unit action, respectively, related to
the chosen system of measure.   In author's opinion,
this practice is extremely dangerous, as the dimensions
of the terms appearing in any equation cannot be
tested.}.
It is apparent that a slow rolling of the scalar
field, $\rho_{\rm K}\dot{\phi}^2/2\ll\rho_{\rm P}
{\cal V}(\phi)$,
implies a negative pressure via Eq.\,(\ref{eq:pq}).

The quintessence equation of state parameter, via
Eqs.\,(\ref{eq:pq}) and (\ref{eq:roq}), reads:
\begin{equation}
\label{eq:w}
w=\frac{p_q}{c^2\rho_q}=\frac{\rho_{\rm K}\dot{\phi}^2/2-\rho_{\rm P}
{\cal V}(\phi)}{\rho_{\rm K}\dot{\phi}^2/2+\rho_{\rm P}{\cal V}(\phi)}~~;
\end{equation}
where, in general, $w$ changes in time.   In
the special case of a static scalar field,
$\dot{\phi}=0$, $w=-1$, and the quintessence
mimics the effect of a cosmological constant.
In the special case of a steady rolling, $\dot
{\phi}=$const, and time-independent potential,
${\cal V}(\phi)=(1/s)(\rho_{\rm K}/\rho_{\rm P})
(\dot{\phi}/2)$, $s$
real number, Eq.\,(\ref{eq:w}) reduces to:
$w=(s-1)/(s+1)$, where $s=0,$ 1/6, 1/5, 1/4,
1/3, 1/2, 1, yields $w=-1$, $-$5/7, $-$2/3,
$-$3/5, $-$1/2, $-$1/3, 0, respectively, $w$ is also
time-independent, and the quintessence mimics
the effect of an ecsessence (e.g., Iliev and
Shapiro, 2001; Horellou and Berge, 2005), which
is a perfect fluid (e.g., Nunes and Mota, 2006).

The general case, $w=w(t)$, can result from a
changing ratio of quintessence kinetic to
potential energy which, in turn, is owing
to the evolution of the related scalar field
potential (e.g., Caldwell et al., 1998a).
Though time-varying equations of state are
closer to the real situation (e.g., Wetterich,
1995; Amendola, 2000; Battye and Weller, 2003;
Mota and van de Bruck, 2004; Percival, 2005;
Manera and Mota, 2006; Nunes and Mota, 2006;
Maio et al., 2006; Basilakos and Voglis, 2007),
the special case of constant $w$ makes
considerable simplification (e.g., Caldwell
et al., 1998a; Wang and Steinhardt, 1998;
Weinberg and Kamionkowski, 2003; Horellou and
Berge, 2005; Maor and Lahav, 2005).
Qualitatively, many of the effects of general
$w(t)$ models can be predicted by interpolating
between models with constant $w$.   If the
quintessence equation of state varies only
slowly with time, the observational predictions
are well approximated by treating $w(t)=$ const
$=w$ (Wang et al., 2000; Percival, 2005).   More
realistic models where $w=w(t)$ such as 2EXP
(e.g., Nunes and Mota, 2006) and SUGRA (e.g.,
Maio et al., 2006) show that $w=$ const to a good
extent for $z\appgeq10$ or $t\appleq0.5$ Gyr.

In dealing with overdensity central collapse,
time scales no longer than a few hudredths of
Gyr are involved, or redshift substantially
larger than about $z=10$.   Accordingly, $w=$
const may safely be assumed in the current
paper.

The combination of Eqs.\,(\ref{eq:fah}),
(\ref{eq:rhoq}), (\ref{eq:rhom}) and
(\ref{eq:Hrmq}) yields at recombination:
\begin{leftsubeqnarray}
\slabel{eq:Hamqa}
&& \left(\frac{\dot{a}_{hi}}{a_{hi}}\right)^2=H^2(a_{hi})=
\frac{2G}{R_{hi}^3}\left[M_{mh}+M_{qh}(a_{hi})\right]~~; \\
\slabel{eq:Hamqb}
&&M_{qh}(a_h)=\frac{4\pi}3\rho_q(a_h)a_h^3=\frac{4\pi}3\rho_q
a_h^{-3w}~~;
\label{seq:Hamq}
\end{leftsubeqnarray}
where $R_h$ is the radius of a sphere filled by
Hubble flow of mass, $M_{mh}+M_{qh}(a_h)$, and
$M_q$ is formally defined as a
quintessence ``mass'' (Caimmi, 2007).
Let a few limiting but relevant
situations be illustrated with more
detail.

In the special case, $w=0$, $f(a_h)=
-3$, according to Eq.\,(\ref{eq:fah}),
and the effect of quintessence is
equivalent to the presence of additional
matter, as shown by Eq.\,(\ref{eq:Frieda}).
Nothing changes with respect to a CDM
universe where $(\Omega_m)_{\rm CDM}=
\Omega_m+\Omega_q$ (e.g., Caldwell et al.,
1998a).   The quintessence mass is
time-independent via Eq.\,(\ref{eq:Hamqb}).

In the special case, $w=-1/3$, $f(a_h)=
-2$, according to Eq.\,(\ref{eq:fah}),
and the effect of quintessence is
equivalent to the presence of additional
curvature, as shown by Eq.\,(\ref{eq:Frieda}).
Nothing changes with respect to a CDM
universe where $(\Omega_k)_{\rm CDM}=
\Omega_k+\Omega_q$ (e.g., Horellou and
Berge, 2005).   The quintessence mass
scales as $a_h$ via Eq.\,(\ref{eq:Hamqb}).

In the special case, $w=-1$, $f(a_h)=
0$, according to Eq.\,(\ref{eq:fah}),
and the effect of quintessence is
equivalent to the presence of a cosmological
constant, as shown by Eq.\,(\ref{eq:Frieda}).
Nothing changes with respect to a CDM
universe in presence of a cosmological
constant where $\Lambda/(3H^2)=\Omega_q$
(e.g., Horellou and Berge, 2005).   The
quintessence mass scales as $a_h^3$ via
Eq.\,(\ref{eq:Hamqb}).

\subsection{Overdensity expansion and central
collapse: general ideas}
\label{epcc}

The dynamics of nonlinear structure
formation in the universe may show a
distinct signature associated to the
nature of the dark energy and a particular
model (Mota and van de Bruck, 2004).
In this view, the behaviour of quintessence
during the nonlinear regime of structure
formation, is conceived as lying between
two limiting cases, namely (i) full
clustering i.e. the scalar field responds
to the infall in the same way as the matter,
and (ii) unclustering i.e. the scalar field
remains homogeneous and the sole effect is
a tidal potential acting on the matter
overdensity.

The general case of partial clustering
can be taken into consideration, at the
expense of a much more complicated
continuity equation for the quintessence
overdensity (e.g., Mota and van de Bruck,
2004; Maor and Lahav, 2005; Nunes and Mota,
2006).   Even though it has been shown
that quintessence cannot be perfectly
smooth (Caldwell et al., 1998a,b),
clustering is usually assumed to be
negligible on scales less than about
100 Mpc (e.g., Wang and Steinhardt,
1998; Weinberg and Kamionkowski, 2003;
Battye and Weller, 2003; Horellou and
Berge, 2005).   It is therefore common
practice to keep the quintessence
homogeneous during the evolution of
overdensities.   The effects of relaxing
this assumption were explored in recent
attempts (e.g., Mota and van de Bruck,
2004; Percival, 2005; Manera and Mota,
2006; Nunes and Mota, 2006).

Given a spherical-symmetric overdensity
and an infinitely thin spherical shell
at a distance, $r$, from the centre,
in the limit of fully clustered
quintessence, the equation of motion
is still expressed by Eq.\,(\ref{eq:Hrmq})
provided the Hubble parameter, $H$, is
replaced by $H_r=\dot{r}/r$ (e.g.,
Nunes and Mota, 2006).   In general,
the evolution of the density of the
scalar field (continuity equation)
within the shell is (e.g., Mota and
van de Bruck, 2004; Maor and Lahav, 2005;
Nunes and Mota, 2006):
\begin{equation}
\label{eq:eco}
\dot{\rho}_q+3\frac{\dot{r}}r(1+w)\rho_q=\Gamma~~;
\end{equation}
where $\Gamma$ describes the quintessence
density change due to the cosmic expansion.
In the limit of fully clustered quintessence,
$\Gamma=0$ as for matter.   In the limit of
unclustered quintessence, $\Gamma=-3(\dot{a}
_h/a_h-\dot{r}/r)(1+w)\rho_q$, and Eq.\,(\ref
{eq:eco}) reduces to its counterpart related
to the quintessence Hubble flow.

The shell expansion is related to cosmic
expansion and overdensity change, as (e.g.,
Nunes and Mota, 2006):
\begin{equation}
\label{eq:aahd}
\frac{\dot{r}}r=\frac{\dot{a}_h}{a_h}-\frac13\frac
{\dot{\delta}(r)}{1+\delta(r)}~~;
\end{equation}
where $\delta(r)$ is the shell overdensity,
defined by Eq.\,(\ref{eq:delRr}).

The real situation may safely be expected
to lie between the above mentioned limiting
cases, where the quintessence is clustering
together with the matter and remains
homogeneous, respectively (e.g., Maor and
Lahav, 2005; Caimmi, 2007).   In general,
overdensities reach turnaround and collapse
earlier in QCDM models with larger
quintessence equation of state parameter,
$w$, and vice versa (e.g., Weinberg and
Kamionkowski, 2003; Horellou and Berge,
2005).   On the other hand, the difference
is small in early times, where the effect
of the dark energy is still negligible
(e.g., Weinberg and Kamionkowski, 2003;
Horellou and Berge, 2005; Nunes and Mota,
2006).

For this reason, the quintessence
equation of state parameter, $w$,
shall be kept constant during overdensity
evolution up to the end of central collapse,
which is expected to occur in early times.
The above condition, in turn, implies that
different evolutions related to different
degrees of quintessence are essentially
indistinguishable except for the later
stages when quintessence starts to dominate
at low redshifts.   In general, the evolution
depends on the quintessence equation of state,
the potential of the scalar field, and the
quintessence contribution to the total energy
budget of the universe at high redshifts.
For further details, refer to e.g., Nunes
and Mota (2006).   Accordingly, the
quintessence within overdensities shall be
assumed as fully clustered to simplify the
calculations.

Under the above mentioned restrictions,
the density ratio of clustered to unclustered
quintessence may be expressed in terms of the
related overdensity value, as (e.g., Nunes
and Mota, 2006):
\begin{equation}
\label{eq:rqrh}
\frac{\rho_q(r)}{\rho_{qh}(a_h)}=\left[\frac{1+\delta(r)}{1+\delta(r_i)}
\right]^{1+w}~~;\qquad w={\rm const}~~;\qquad\Gamma=0~~;
\end{equation}
where $\rho_{qh}$ is the density within
the quintessence Hubble flow and the index,
$i$, denotes the beginning of evolution.

The integration of Eq.\,(\ref{eq:eco})
where $w={\rm const}$, $\Gamma=0$, yields (e.g.,
Caimmi, 2007):
\begin{equation}
\label{eq:dMqx}
\frac{\rho_q(r)}{\rho_q(r_{\rm max})}=\left(\frac r{r_{\rm max}}
\right)^{-3(1+w)}~~;\qquad w={\rm const}~~;\qquad\Gamma=0~~;
\end{equation}
with regard to the turnaround configuration,
$r=r_{\rm max}$.   Accordingly, the
quintessence density is related to the
radius by a power law with
exponent, $b-3=-3w-3$.   The special
case, $w=-1/3$, $b=1$, corresponds to
an isothermal sphere; $w=-1$, $b=3$,
to a homogeneous sphere; $w=0$, $b=0$,
to a Roche sphere.

The quintessence mass, $M_q(r)$, enclosed
by a homogeneous spherical overdensity
with equal density and radius as in
the infinitely thin sperical shell
under consideration, scales as:
\begin{equation}
\label{eq:Mqx}
\frac{M_q(r)}{M_q(r_{\rm max})}=\left(\frac r{r_{\rm max}}
\right)^{-3w}~~;
\end{equation}
which makes the mass increase as the radius
increases, and vice versa.

The validity of Newton's theorem and
McLaurin's theorem being independent
of the value of the gravitation constant,
$G$, homogeneous overdensities where the
quintessence is fully clustered evolve
in the same way as within static
quintessence density profiles, defined
by Eqs.\,(\ref{eq:dMqx}) or (\ref{eq:Mqx}),
even if the value of the interaction
strength is different.   In this view, $M_q(r)$
is the quintessence mass related to the
static profile, enclosed within the
volume of the homogeneous overdensity
of radius, $r$.

Owing to the above mentioned theorems,
(i) given a quintessence spherical corona,
related to the static profile, the
gravitational action on a selected point
enclosed by the inner surface of the
corona, is null and (ii) given a
quintessence sphere related to the
static profile, enclosing a mass,
$M_q(r^\prime)$, the gravitational action
on a selected point outside the
surface is equivalent to its counterpart
exerted by a central point of equal
mass, $M_q(r^\prime)$.

Using the above results, the evolution
of overdensity infinitely thin, shperical
shells may be determined in the limit of
fully clustered quintessence.   To this
aim, the gravitational potential induced
by the quintessence is conceived as
arising from a distribution where any
two ``particles'', idealized as ``mass
points'', $m_{qi}$ and $m_{qj}$, interact
with strength, $(1+3w)G$, according to a
Newton-like law, $F_{ij}=(1+3w)Gm_{qi}m_
{qj}/r_{ij}$.   Then the results related
to two-component matter distributions
(e.g., Limber, 1959; Brosche et al., 1983;
Caimmi et al., 1984; Caimmi and Secco, 1992)
may be generalized to the case, where a
subsystem is made of quintessence.
For further details, refer to Caimmi
(2007).

In the special case under consideration
$(w={\rm const},~\Gamma=0)$ the expansion of
an infinitely thin spherical shell
enclosing a matter mass, $M_m(r)$,
and a quintessence mass, $M_q(r)$,
within a radius, $r$, is expressed
as (e.g., Mota and van de Bruck, 2004):
\begin{equation}
\label{eq:expa}
\ddot{r}+\frac{GM_m(r)}{r^2}+\frac{(1+3w)GM_q(r)}{r^2}=0~~;
\end{equation}
where mass conservation holds for
matter, $M_{\rm m}(r)=M_m(r_i)$, and
quintessence mass changes in time
according to Eq.\,(\ref{eq:Mqx}).
The special cases, $w=0$ and $w=-
1/3$, relate to flat $(\Omega_{\rm CDM}=
\Omega_m+\Omega_q=1)$ and open
$(\Omega_{\rm CDM}=\Omega_m=1-\Omega_q)$
CDM universes.

The quintessence mass, $M_q(r)$,
is assumed to change in
time according to Eq.\,(\ref
{eq:Mqx}), which is related to the
quintessence mass enclosed by
a homogeneous spherical overdensity
with equal density and radius as in
the infinitely thin shell under
consideration.   The above approximation
overstimates the collapse rate, where
the larger effect corresponds to the
lower quintessence equation of state
parameter, and vice versa.   In any
case, the effect of the dark energy
is negligible in early times
(e.g., Weinberg and Kamionkowski, 2003;
Horellou and Berge, 2005; Nunes and Mota,
2006), when central collapse is
completed.
                             
Following a
similar procedure as in presence of
sole matter (e.g., Caimmi, 1989), an
integration of Eq.\,(\ref{eq:expa})
yields:
\begin{lefteqnarray}
\label{eq:expv}
&& \dot{r}=\mp\left\{\dot{r_i}^2-\frac{2GM_m(r_i)}{r_i}+
\frac{2GM_m(r_i)}r\right. \nonumber \\
&& \left.-\frac{2(1+3w)GM_q(r_{\rm max})(r_{\rm max})^{3w}}
{r_i^{1+3w}}+\frac{2(1+3w)GM_q(r_{\rm max})(r_{\rm max})^{3w}}
{r^{1+3w}}\right\}^{1/2};
\end{lefteqnarray}
where the plus is related to expansion,
and the minus to contraction.

To a good extent, at recombination epoch
overdensities still expand together with
the universe, which implies the boundary
conditions:
\begin{equation}
\label{eq:cir}
r_i=r_{hi}~~;\qquad\dot{r}_i=\dot{r}_{hi}~~;\qquad
r_{hi}=a_{hi}r_{h0}~~;
\end{equation}
where $r_h$ is the radius of a sphere
with zero overdensity.

The particularization of Eq.\,(\ref{eq:Hrmq})
to $a_h=a_{hi}$, after combination with
Eqs.\,(\ref{eq:M}), (\ref{eq:fah}),
(\ref{eq:rhoq}), (\ref{eq:rhom}), (\ref
{eq:Mqx}), and (\ref{eq:cir}) yields:
\begin{equation}
\label{eq:rpi}
\dot{r}_i=\left[\frac{2GM_m(r_i)}{(1+\overline{\delta}_i)r_i}+
\frac{2GM_q(r_{\rm max})(r_{\rm max})^{3w}}{r_i^{1+3w}}
\right]^{1/2}~~;
\end{equation}
provided isotropic radial motions are
dominant.

At turnaround, $\dot{r}=0$ by definition,
which implies by use of Eq.\,(\ref{eq:expv})
particularized to $r=r_{\rm max}$:
\begin{lefteqnarray}
\label{eq:exp0}
&& \dot{r}_i^2-\frac{2GM_m(r_i)}{r_i}-\frac{2(1+3w)GM_q(r_{\rm max})
(r_{\rm max})^{3w}}{r_i^{1+3w}} \nonumber \\
&& =-\frac{2GM_m(r_i)}{r_{\rm max}}-\frac{2(1+3w)GM_q(r_{\rm max})}
{r_{\rm max}}~~;
\end{lefteqnarray}
and the substitution of $\dot{r}_i^2$ into
the general expression, Eq.\,(\ref{eq:expv}),
yields:
\begin{lefteqnarray}
\label{eq:exvm}
&& \dot{r}=\mp\left[\frac{2GM_m(r_i)}r-
\frac{2GM_m(r_i)}{r_{\rm max}}\right. \nonumber \\
&& \left.+\frac{2(1+3w)GM_q(r_{\rm max})(r_{\rm max})^{3w}}
{r^{1+3w}}-\frac{2(1+3w)GM_q(r_{\rm max})}{r_{\rm max}}
\right]^{1/2}~~;
\end{lefteqnarray}
where the turnaround radius, $r_{\rm max}$,
has replaced the recombination radius, $r_i$.

Using the dimensionless variables:
\begin{leftsubeqnarray}
\slabel{eq:alta}
&& \alpha=\frac r{r_{\rm max}}~~;\qquad\tau=\frac t{t_{\rm max}^
\dagger}~~; \\
\slabel{eq:altb}
&& t_{\rm max}^\dagger=\left[\frac{8GM_m(r_i)}{\pi^2
(r_{\rm max})^3}\right]^{-1/2}=\left[\frac{32}{3\pi}G\overline
{\rho}(r_{\rm max})\right]^{-1/2}~~;
\label{seq:alt}
\end{leftsubeqnarray}
together with the fractional mass:
\begin{leftsubeqnarray}
\slabel{eq:ma}
&& m=m(r_{\rm max})=\frac{M_q(r_{\rm max})}{M_m(r_i)}~~; \\
\slabel{eq:mb}
&& \frac1{1+m}=\frac{M_m(r_i)}{M_{\rm tot}(r_{\rm max})};~
\frac m{1+m}=\frac{M_q(r_{\rm max})}{M_{\rm tot}(r_{\rm max})};
~M_{\rm tot}=M_m+M_q;
\label{seq:m}
\end{leftsubeqnarray}
allow to cast Eq.\,(\ref{eq:exvm})
under the equivalent form:
\begin{equation}
\label{eq:exam}
\frac{\diff\alpha}{\diff\tau}=\mp\frac\pi2\left[\left(\frac1
\alpha-1\right)+(1+3w)m\left(\frac1{\alpha^{1+3w}}-1\right)
\right]^{1/2}~~;
\end{equation}
where the special cases, $w=0$ and
$w=-1/3$, reproduce the results related
to flat $(\Omega_{\rm CDM}=\Omega_m+\Omega_
q=1)$ and open $(\Omega_{\rm CDM}=\Omega_m=
1-\Omega_q)$ CDM universes (e.g., Ryden
and Gunn, 1987; Caimmi, 1989).

It is worth noticing the scaling time,
$t_{\rm max}^\dagger$, defined by
Eq.\,(\ref{eq:altb}), coincides with
the actual matter overdensity free-fall
time only for CDM universes i.e. $M_
{\rm tot}(r_{\rm max})=M_m(r_{\rm max})$,
$M_q(r_{\rm max})=0$.   In the case under
discussion,
overdensity collapse is equivalent to
pure matter collapse within a static
quintessence density profile, $\rho_q
(r)/\rho_q(r_{\rm max})=(r/r_{\rm max})^
{-3(1+w)}$, according to Eq.\,(\ref
{eq:dMqx}).   The related free-fall
time, to be calculated numerically,
is the overdensity free-fall time.
Then the dimensionless time, $\tau$,
is different from unity at turnaround,
unless CDM universes are considered
(e.g., Ryden and Gunn, 1987; Caimmi,
1989).   In any case, CDM models
make a useful reference case, and
for this reason the main results shall
be recalled in the following.

In the special case, $w=0$, the
effect of quintessence is equivalent
to the effect of additional matter,
$\Omega_{\rm CDM}=\Omega_m+\Omega_q=1$,
and after a re-definition of the
dimensionless time, Eq.\,(\ref{eq:exam})
reduces to:
\begin{leftsubeqnarray}
\slabel{eq:exw0a}
&& \frac{\diff\alpha}{\diff\tau}=\mp\frac\pi2\frac{\sqrt{\alpha
-\alpha^2}}\alpha~~; \\
\slabel{eq:exw0b}
&& \tau=\sqrt{1+m}\,\frac t{t_{\rm max}^\dagger}~~;
\label{seq:exw0}
\end{leftsubeqnarray}
where the scaling time, $t_{\rm max}^
\dagger/\sqrt{1+m}$, represents the
overdensity free-fall time with respect
to the turnaround configuration.   An
integration yields (e.g., Caimmi, 1989):
\begin{equation}
\label{eq:alw0}
\tau-\tau_i=\mp\frac2\pi\left[-\sqrt{\alpha-\alpha^2}+
\sqrt{\alpha_i-\alpha_i^2}-\arcsin\sqrt{1-\alpha}+
\arcsin\sqrt{1-\alpha_i}\right]~~;
\end{equation}
in the idealized situation where radial
motion is preserved, $\alpha=1$ and
$\tau=2i-1$ at $i$-th turnaround;
$\alpha=0$ and $\tau=2i$ at $i$-th
point-like configuration; and
Eq.\,(\ref{eq:alw0}) can be splitted as:
\begin{leftsubeqnarray}
\slabel{eq:aleca}
&& \tau=(2i-1)-\frac2\pi\left[\sqrt{\alpha
-\alpha^2}+\arcsin\sqrt{1-\alpha}\right]~~; \\
\slabel{eq:alecb}
&& \tau=(2i-1)+\frac2\pi\left[\sqrt{\alpha
-\alpha^2}+\arcsin\sqrt{1-\alpha}\right]~~;
\label{seq:alec}
\end{leftsubeqnarray}
which describe $i$-th expansion and $i$-th
contraction, respectively.

For sake of completeness, it is worth
recalling the parametric equations (e.g.,
Ryden and Gunn, 1987; Caimmi, 1989):
\begin{leftsubeqnarray}
\slabel{eq:para}
&& \tau=\frac{\theta-\sin\theta}\pi~~;\qquad\alpha=
\frac{1-\cos\theta}2~~; \\
\slabel{eq:parb}
&& \frac{\diff\tau}{\diff\theta}=\frac2\pi\alpha~~;\qquad
\frac{\diff\alpha}{\diff\theta}=\mp\sqrt{\alpha-\alpha^2}~~;
\label{seq:par}
\end{leftsubeqnarray}
where Eq.\,(\ref{seq:exw0}) has been used
for the expression of $\diff\alpha/\diff
\theta$.

In the special case, $w=-1/3$, the effect
of quintessence is equivalent to the
effect of additional curvature, $\Omega_
{\rm CDM}=\Omega_m=1-\Omega_q$, and Eq.\,(\ref
{eq:exam}) reduces to:
\begin{equation}
\label{eq:exw1}
\frac{\diff\alpha}{\diff\tau}=\mp\frac\pi2\frac
{\sqrt{\alpha-\alpha^2}}\alpha~~;
\end{equation}
which coincides with Eq.\,(\ref{eq:exw0a}),
and the validity of Eqs.\,(\ref{eq:alw0})-(\ref{seq:par})
is maintained.

In the special case, $w=-1$, the effect of
quintessence is equivalent to the effect of
a cosmological constant, $\Lambda/(3H^2)=
\Omega_q$, and Eq.\,(\ref{eq:exam}) reduces to:
\begin{equation}
\label{eq:exwc}
\frac{\diff\alpha}{\diff\tau}=\mp\frac\pi2\left[\left(\frac1
\alpha-1\right)-2m\left(\alpha^2-1\right)\right]^{1/2}~~;
\end{equation}
which has no analytical solution and must
be solved via numerical integration.

\subsection{Overdensity expansion and central collapse:
a special case}
\label{ecsc}

To the aim of the current attempt, only the
early phase of overdensity evolution is
considered, from recombination to the end
of central collapse, which coincides with
the onset of shell crossing in the limit of
classical mechanics, where black hole formation
does not occur and radial motions are
reversed off from the centre.    The central
collapse may safely be expected to end in a
considerably short time, even for massive
overdensities.   Accordingly, the quintessence
effect may be neglected to a first extent.

From this point on, further effort shall be
devoted to the special case, $w=-1/3$, for
the following reasons: (i) the overdensity
equation of motion can analytically be
integrated, according to Eq.\,(\ref{eq:alw0}),
even in presence of (fully clustered)
quintessence; (ii) the (matter + quintessence)
overdensity behaves as a pure matter overdensity
within an open CDM universe where $\Omega_{\rm CDM}=
\Omega_m=1-\Omega_q$ (e.g., Horellou and Berge,
2005), and the related results may be extended
to the case under discussion; (iii) overdensities
reach turnaround and collapse earlier for
increasing values of the quintessence equation
of state parameter, $w$, and vice versa (e.g.,
Horellou and Berge, 2005), which makes the case
under discussion, $w=-1/3$, a lower limit with
regard to the range of interest, $-1\le w\le-
1/3$, where the maximum difference does not
exceed about 15\% if collapse ends at present
for $w=-1$ (Horellou and Berge, 2005).
Concerning central collapse, the above
difference is expected to be higly reduced.

According to Eq.\,(\ref{eq:exw1}), at
turnaround $\alpha=1$, and the scaling time,
$t_{\rm max}^\dagger$, coincides with the
free-fall time of the matter subsystem,
which implies $\tau_{\rm max}=1$.   For
the initial configuration, the combination
of Eqs.\,(\ref{seq:Rmax}), (\ref{eq:alta}), and
(\ref{eq:aleca}) yields:
\begin{leftsubeqnarray}
\slabel{eq:coia}
&& \alpha_i=\Delta_i=1-(\Omega_m)_i^{-1}(1+\overline{\delta}_i)^{-1}
\approx\frac{\overline{\delta}_i}{1+\overline{\delta}_i};~
\left\vert1-(\Omega_m)_i^{-1}\right\vert\ll\overline{\delta}_i\ll1; \\
\slabel{eq:coib}
&& \tau_i=1-\frac2\pi\left[\sqrt{\Delta_i-\Delta_i^2}-
\arcsin\sqrt{1-\Delta_i}\right]~~.
\label{seq:coi}
\end{leftsubeqnarray}

Using the definition of mean density and
matter mass conservation, the free-fall
time, $t_{\rm max}$, via Eq.\,(\ref{eq:altb})
takes the expression:
\begin{equation}
\label{eq:tmax}
t_{\rm max}=t_{\rm max}^\dagger=\left[\frac{32}{3\pi}G
(\overline{\rho}_m)_i\right]^{-1/2}\alpha_i^{-3/2}~~;
\end{equation}
where $(\overline{\rho}_m)_i$ is the mean
density within a sphere of radius, $r_i$,
centered on a local maximum at recombination
epoch.   The combination of Eqs.\,(\ref{eq:delRr}),
(\ref{eq:coia}), and (\ref{eq:tmax}) yields:
\begin{equation}
\label{eq:tmhd}
t_{\rm max}(r_i)=\left[\frac{32}{3\pi}G(\rho_{hm})_i\right]^
{-1/2}\frac{1+\overline{\delta}_i(r_i)}{[\Delta_i(r_i)]^{3/2}}~~;
\end{equation}
where $(\rho_{hm})_i$ is the density of
the matter Hubble flow at recombination
epoch.

The mean overdensity, $\overline{\delta}$,
can be related to the initial value,
$\overline{\delta}_i$, using mass conservation
together with Eqs.\,(\ref{eq:M}), (\ref{seq:alec}),
and (\ref{eq:exwc}).   The result is:
\begin{leftsubeqnarray}
\slabel{eq:mova}
&& \frac{1+\overline{\delta}}{1+\overline{\delta}_i}=
\left(\frac\alpha{\alpha_h}\right)^3~~; \\
\slabel{eq:movb}
&& \alpha_h=\frac{r_h}{r_{\rm max}}~~;
\label{seq:mov}
\end{leftsubeqnarray}
where $r_h$ is the radius of a sphere with
zero overdensity.

Starting from the definition of local
density within an infinitely thin
spherical shell, and mean density
within the volume enclosed by the
shell:
\begin{lefteqnarray}
\label{eq:lode}
&& \rho_m(r)=\frac3{4\pi r^2}\frac{\diff M_m}{\diff r}~~; \\
\label{eq:mede}
&& \overline{\rho}_m(r)=\frac3{4\pi}\frac{M_m(r)}{r^3}~~;
\end{lefteqnarray}
mass conservation within the shell and the
volume bounded by the shell, with regard
to the initial configuration, reads (e.g.,
Peebles, 1980, Chap.\,II, \S\,19C; Ryden
and Gunn, 1987; Caimmi, 1990):
\begin{lefteqnarray}
\label{eq:macl}
&& \rho_m(r)=\rho_m(r_i)\left(\frac r{r_i}\right)^{-2}
\left(\frac{\partial r}{\partial r_i}\right)^{-1}~~; \\
\label{eq:macm}
&& \overline{\rho}_m(r)=\overline{\rho}_m(r_i)\left(\frac
r{r_i}\right)^{-3}~~;
\end{lefteqnarray}
and the combination of Eqs.\,(\ref{eq:M})
and (\ref{seq:mov})-(\ref{eq:macm}) yields:
\begin{leftsubeqnarray}
\slabel{eq:gda}
&& \frac{\rho_m(r)}{\overline{\rho}_m(r)}=\frac{1+\delta(r)}
{1+\overline{\delta}(r)}=\frac{1+\delta_i(r_i)}{1+\overline
{\delta}_i(r_i)}g(r)~~; \\
\slabel{eq:gdb}
&& g(r)=\frac r{r_i}\left(\frac{\partial r}{\partial r_i}\right)^{-1}~~;
\label{seq:gd}
\end{leftsubeqnarray}
where the mean overdensity, $\overline
{\delta}(r)$, can be determined for a
fixed cosmological model via Eqs.\,(\ref
{eq:alw0}) and (\ref{seq:mov}), and the
explicit expression of the partial derivative,
$\partial r/\partial r_i$, allows the
explicit expression of the local
overdensity, $\delta(r)$, via Eqs.\,(\ref
{seq:mov}).

The related procedure, involving long but
stimulating algebra, has been performed
in different situations related to CDM
cosmologies: overdensity evolution in
Einstein-de Sitter universes (Peebles,
1980, Chap.\,II, \S\,19C; Ryden and Gunn,
1987; Caimmi, 1989); overdensity evolution
in Einstein-de Sitter universes with the
effects of acquisition of angular
momentum on the expansion included
(Caimmi, 1990); overdensity evolution
in high-energy $(\Omega_m\appgeq0.1)$
universes (Andriani and Caimmi, 1991);
overdensity evolution in high-energy
$(\Omega_m\appgeq0.1)$ universes with the
effects of acquisition of angular
momentum on the expansion included
(Andriani and Caimmi, 1994).   The
third above mentioned case (Andriani
and Caimmi, 1991) shall be considered
in the current attempt.

Accordingly, the explicit expression
of Eq.\,(\ref{eq:gdb}) reads:
\begin{lefteqnarray}
\label{eq:ges}
&& [g(r)]^{-1}=1+3\frac{1-\Delta_i}{\Delta_i}\frac{\overline{\delta}_i}
{1+\overline{\delta}_i}\left(1-\frac{\delta_i}{\overline{\delta}_i}
\right)\left\{1-\frac{\Delta_i^2}{\sqrt{\Delta_i-\Delta_i^2}}\frac
{\mp\sqrt{\alpha-\alpha^2}}{\alpha^2}\times\right. \nonumber \\
&& 
\left.\times\left[1+\frac{3\pi}4(1+\overline{\delta}_i)^{1/2}\left(
\frac{1-\Delta_i}{\Delta_i^3}\right)^{1/2}\left(
1+\frac13\frac{\Delta_i}{1-\Delta_i}\right)(\tau-\tau_i)\right]\right\}~~;
\end{lefteqnarray}
where the positive and the negative
sign within parentheses are related
to shell expansion and contraction,
respectively. For a formal derivation,
see Appendix \ref{a:g}.

For density profiles obeying Eq.\,(\ref
{eq:ropi}),
overdensity central collapse is completed
at a dimensionless time, $\tau(r_{\rm pk})
=2$, according to Eq.\,(\ref{eq:alecb}),
where $r_{\rm pk}$ is the radius of
the homogeneous peak at recombination
epoch.

Let $\tau_c(r_i)$ be the dimensionless
time related to an infinitely thin
spherical shell of initial radius, 
$r_i$, at the end of
central collapse i.e. $t=2t_{\rm m
ax}(r_{\rm pk})$.  The combination
of Eqs.\,(\ref{seq:alt}) and (\ref
{eq:tmhd}) produces:
\begin{equation}
\label{eq:tauc}
\tau_c(r_i)=2\frac{1+\delta_{\rm pk}}{1+\overline{\delta}_i(r_i)}
\left[\frac{\overline{\delta}_i(r_i)}{\delta_{\rm pk}}\right]^
{3/2}~~;
\end{equation}
where $\delta_{\rm pk}=\overline
{\delta}_i(r_{\rm pk})$ is the
overdensity value of the homogeneous
peak, and $\overline{\delta}_i(r_i)$
is related to the initial density
profile via Eqs.\,(\ref{eq:derm})
and (\ref{eq:deRm}).

The related dimensionless distance,
$\alpha_c(r_i)$, can be determined as
the solution to the transcendental
equation deduced from Eq.\,(\ref
{eq:alecb}):
\begin{lefteqnarray}
\label{eq:alct}
&& \tau_c(r_i)=1+\frac2\pi\left[\sqrt{\alpha_c(r_i)-\alpha_c^2(r_i)}+
\arcsin\sqrt{1-\alpha_c(r_i)}\right]~~;
\end{lefteqnarray}
and the combination of Eqs.\,(\ref
{eq:alta}) and (\ref{eq:coia}) yields:
\begin{equation}
\label{eq:alc}
\alpha_c(r_i)=\frac{r_c(r_i)}{r_i}\frac{r_i}{r_{\rm max}(r_i)}=
\frac{r_c(r_i)}{r_i}\alpha_i(r_i)=\frac{r_c(r_i)}{r_i}\Delta_i(r_i)~~;
\end{equation}
which allows an explicit expression
of the radius of an infinitely thin
spherical shell at the end of central
collapse, as:
\begin{equation}
\label{eq:rcri}
r_c(r_i)=\frac{\alpha_c(r_i)r_i}{\Delta_i(r_i)}~~;
\end{equation}
where $r_i$ is the initial radius at
recombination epoch.

The density profile at the end of
central collapse, follows from the
particularization of Eqs.\,(\ref
{seq:gd}) and (\ref{eq:ges}) to
the case of interest.   The result
is:
\begin{leftsubeqnarray}
\slabel{eq:delga}
&& \frac{\rho(r_c)}{\overline{\rho}(r_c)}=\frac{1+\delta(r_c)}
{1+\overline{\delta}(r_c)}=\frac{1+\delta_i(r_i)}{1+\overline
{\delta}_i(r_i)}g(r_c)~~; \\
\slabel{eq:delgb}
&& [g(r_c)]^{-1}=1+3\frac{1-\Delta_i(r_i)}{\Delta_i(r_i)}\frac{\overline
{\delta}_i(r_i)}{1+\overline{\delta}_i(r_i)}\left[1-\frac{\delta_i(r_i)}
{\overline{\delta}_i(r_i)}\right]\left\{1-\frac{[\Delta_i(r_i)]^2}
{\sqrt{\Delta_i(r_i)-[\Delta_i(r_i)]^2}}\right. \nonumber \\
&& \phantom{[g(r_c)]^{-1}=}\left.\times
\frac{\mp\sqrt{\alpha_c(r_i)-
[\alpha_c(r_i)]^2}}{[\alpha_c(r_i)]^2}
\left[1+\frac{3\pi}4[1+\overline{\delta}_i(r_i)]^{1/2}\left[
\frac{1-\Delta_i(r_i)}{\Delta_i^3(r_i)}\right]^{1/2}\right.\right.
\nonumber \\
&& \phantom{[g(r_c)]^{-1}=}\left.\left.\times
\left[1+\frac13\frac{\Delta_i(r_i)}{1-\Delta_i(r_i)}\right]
[\tau_c(r_i)-\tau_i(r_i)]\right]\right\}~~;
\label{seq:delg}
\end{leftsubeqnarray}
where global and local overdensities,
$\overline{\delta}_i$ and $\delta_i$,
are determined by the knowledge of the
initial density profile at recombination
epoch, via Eqs.\,(\ref{eq:derm}) and
(\ref{eq:der}), respectively.
                       
The overdensity evolution obeying an
initial density profile defined by
Eq.\,(\ref{eq:ropi}), lies between
the limiting situations, $b=0$ and
$b=3$.   The special case, $b=0$,
corresponds to a Roche sphere i.e.
a mass point surrounded by a massless
atmosphere.   Accordingly, $r_{\rm pk}=0$
and the hole is present from the
beginning of evolution.   The special
case, $b=3$, corresponds to a homogeneous
sphere which collapses into a hole
after a time, $t=2t_{\rm max}(R_i)$.
In the general case, $0<b<3$, the
homogeneous peak collapses into a
hole after a time, $t=2t_{\rm max}
(r_{\rm pk})$, while the underlying
envelope virializes due to deviations
from radial motion.

According to the results of Subsect.\,\ref
{idep}, the dimensionless radius,
$\Xi=R/r_{\rm pk}$, Eq.\,(\ref{seq:Rrc}),
and the dimensionless densities, $\overline
{\rho}_{\rm tr}/\rho_{\rm pk}$ and $\rho_
{\rm tr}/\rho_{\rm pk}$, Eqs.\,(\ref{eq:romc})
and (\ref{eq:roRc}), respectively, depend
only on the input parameters, $b$ and
$\kappa$.   For changing masses and peak
heights, the above mentioned quantities
necessarily remain fixed.   With regard
to the mass ratio, $\kappa=M_{\rm pk}/M_
{\rm tr}$, an empirical relation between
hole and hosting spheroid mass (e.g.,
Ferrarese and Ford, 2005) cannot be used
for for the following reasons.

First, the above mentioned relation
holds for ellipticals and early-type
spirals, but the disk mass is not
taken into consideration.   On the
other hand, the disk is dominant in
late-type spirals, and both the
proto-bulge and the proto-disk,
together with the (nonbaryonic)
dark matter, must be included in
the initial density profile at
recombination epoch.

Second, the dark matter is dominant
with respect to the baryonic matter,
and the relation between dark matter
and hole mass seems to be nonlinear,
as:
\begin{equation}
\label{eq:Mhd}
\frac{M_{\rm hole}}{M_{\rm dark}}=5.8\,10^{-5}\left(\frac1{100}\frac
{M_{\rm dark}}{{\rm M}_{10}}\right)^\beta\frac{1+z}7~~;\qquad10\appleq M_
{\rm dark}/{\rm M}_{10}\appleq1000~~;
\end{equation}
where $z$ is the redshift and $\beta$
an exponent to be fixed.   Special
values are: $(\beta,z)=(0.65,6/29)$
(Ferrarese, 2002); $(\beta,z)=
(0.27,19/58)$ (Baes et al., 2003);
and $\beta=0.39$
(Shankar and Mathur, 2007).   In
addition, Eq.\,(\ref{eq:Mhd}) is
close to its counterpart calibrated
locally through statistical arguments
(Shankar et al., 2006) and to its
counterparts obtained using
semianalytical models (Granato et
al., 2004; Lapi et al., 2006).
It can be seen that a redshift-dependent
superlinear relation, $M_{\rm hole}/M_
{\rm dark}\propto M_{\rm dark}^\beta$,
$0.3\appleq\beta\appleq0.7$, is best
suited to represent the luminosity
function in active galactic nuclei
(Shankar et al., 2007).   Accordingly,
Eq.\,(\ref{eq:Mhd}) shall be used in
computations.
An extrapolation to the mass range
under consideration, $10^{-1}\le M/
{\rm M}_{10}\le10^6$, via Eq.\,(\ref{eq:Mhd})
yields $10^{-8}<M_{\rm hole}/{\rm M}_{10}<
10^4$ in the special case, $\beta=0.65$,
$z=6/29$.

If the lack of low-mass $(10^{-7}\appleq M_
{\rm hole}/{\rm M}_{10}\appleq10^{-4})$ holes is
real, a possible interpretation could be
the following.   Secondary perturbations
naturally arise around a local density maximum
(e.g., Ryden and Gunn, 1987) and virialize
when the hosting overdensity is still expanding.
For total clump
number or typical clump mass independent of
overdensity mass, it can be seen that the
distance between nearest clumps (provided
they are uniformly distributed within the
overdensity volume) increases as the
overdensity mass increases, and vice versa.
Accordingly, tidal interactions and
related departure from radial motions are
expected to be more efficient in low-mass
$(M\appleq10^2{\rm M}_{10})$ overdensities,
which could prevent hole formation. 


\section{Results}
\label{resu}

Computations have been performed up to
the end of the central collapse, within
the mass range, $-1\le\log(M/{\rm M}_{10})\le
6$, for mean overdensity heights, $\overline
{\nu}=1,2,3,4,$ and an assumed quintessence
equation of state parameter, $w=-1/3$.
The density profile at the end of central
collapse, and related values of dimensionless
times, $\tau_c(r_i)$, and dimensionless radii,
$\alpha_c(r_i)$, are weakly dependent on the
mean overdensity height, to a major extent as the
initial shell radius, $r_i$, is selected near
the truncation radius.   In addition,
different overdensity heights related to low masses
exhibit slightly larger differences with
respect to their counterparts related to
high masses.   Also, differences in $\tau_c$
and $\alpha_c$ are more pronounced (but
still negligible) in comparison to
differences in density, provided the
remaining parameters are kept fixed.

The ratio of homogeneous peak radius to
truncation radius, $\Xi^{-1}=r_{\rm pk}/
R_{\rm tr}$, mean density to homogeneous peak
density, $\overline{\rho}_{\rm tr}/\rho_{\rm pk}$,
and local density at truncation radius
to homogeneous peak density, $\rho_{\rm
tr}/\rho_{\rm pk}$, depend only on the
slope of the external density profile,
$b$, and the ratio of homogeneous peak
mass to total mass within the truncation
radius, $\kappa=M_{\rm pk}/M_{\rm tr}$,
according to Eqs.\,(\ref{seq:Rrc}),
(\ref{eq:romc}), and (\ref{eq:roRc}),
respectively.   The dependence within
the ranges, $2.85\le b\le3$, $-8\le
\log\kappa\le-2$, is shown in Tab.\,\ref
{t:kab}.
\begin{table}
\begin{tabular}{lllllll}
\hline
\hline
\multicolumn{1}{c}{$\log\kappa=-8$} &
\multicolumn{1}{c}{$-7$}
&
\multicolumn{1}{c}{$-6$} &
\multicolumn{1}{c}{$-5$} &
\multicolumn{1}{c}{$-4$} &
\multicolumn{1}{c}{$-3$} &
\multicolumn{1}{c}{$-2$} \\
\hline
1.588 E$-$3 & 3.562 E$-$3 & 7.990 E$-$3 & 1.792 E$-$2 & 4.021 E$-$2
& 9.019 E$-$2 & 2.023 E$-$1 \\
4.003 E$-$1 & 4.519 E$-$1 & 5.101 E$-$1 & 5.758 E$-$1 & 6.500 E$-$1
& 7.337 E$-$1 & 8.278 E$-$1 \\
3.803 E$-$1 & 4.293 E$-$1 & 4.846 E$-$1 & 5.470 E$-$1 & 6.175 E$-$1
& 6.971 E$-$1 & 7.869 E$-$1 \\
1.530 E$-$0 & 1.241 E$-$0 & 9.823 E$-$1 & 7.587 E$-$1 & 5.580 E$-$1
& 3.802 E$-$1 & 2.233 E$-$1 \\
      & & & & & & \\
1.764 E$-$3 & 3.902 E$-$3 & 8.632 E$-$3 & 1.910 E$-$2 & 4.224 E$-$2
& 9.345 E$-$2 & 2.067 E$-$1 \\
5.487 E$-$1 & 5.941 E$-$1 & 6.432 E$-$1 & 6.963 E$-$1 & 7.539 E$-$1
& 8.161 E$-$1 & 8.833 E$-$1 \\
5.304 E$-$1 & 5.743 E$-$1 & 6.217 E$-$1 & 6.731 E$-$1 & 7.287 E$-$1
& 7.890 E$-$1 & 8.542 E$-$1 \\
8.455 E$-$1 & 7.046 E$-$1 & 5.745 E$-$1 & 4.543 E$-$1 & 3.433 E$-$1
& 2.408 E$-$1 & 1.465 E$-$1 \\
      & & & & & & \\
1.953 E$-$3 & 4.262 E$-$3 & 9.302 E$-$3 & 2.030 E$-$2 & 4.431 E$-$2
& 9.672 E$-$2 & 2.111 E$-$1 \\
7.444 E$-$1 & 7.741 E$-$1 & 8.049 E$-$1 & 8.369 E$-$1 & 8.702 E$-$1
& 9.048 E$-$1 & 9.407 E$-$1 \\
7.320 E$-$1 & 7.612 E$-$1 & 7.915 E$-$1 & 8.230 E$-$1 & 8.557 E$-$1
& 8.898 E$-$1 & 9.252 E$-$1 \\
3.603 E$-$1 & 3.083 E$-$1 & 2.582 E$-$1 & 2.100 E$-$1 & 1.637 E$-$1
& 1.192 E$-$1 & 7.654 E$-$2 \\
      & & & & & & \\
2.154 E$-$3 & 4.642 E$-$3 & 1.000 E$-$2 & 2.154 E$-$2 & 4.642 E$-$2
& 1.000 E$-$1 & 2.154 E$-$1 \\
1.000 E$-$0 & 1.000 E$-$0 & 1.000 E$-$0 & 1.000 E$-$0 & 1.000 E$-$0
& 1.000 E$-$0 & 1.000 E$-$0 \\
1.000 E$-$0 & 1.000 E$-$0 & 1.000 E$-$0 & 1.000 E$-$0 & 1.000 E$-$0
& 1.000 E$-$0 & 1.000 E$-$0 \\
1.270 E$-$2 & 1.270 E$-$2 & 1.270 E$-$2 & 1.270 E$-$2 & 1.270 E$-$2
& 1.270 E$-$2 & 1.270 E$-$2 \\

\hline\hline
\end{tabular}
\caption{From up to down of each block:
ratio of homogeneous peak radius
to truncation radius, $\Xi^{-1}=r_{\rm pk}
/R_{\rm tr}$ (first lines), mean density
to homogeneous peak density, $\overline
{\rho}_{\rm tr}/\rho_{\rm pk}$ (second lines),
local density at truncation radius to
homogeneous peak density, $\rho_{\rm tr}/
\rho_{\rm pk}$ (third lines),  and peak
overdensity, $\delta_{\rm pk}$,
(fourth lines), for different
choices of ratio of homogeneous peak mass to
total mass within the truncation radius,
$\kappa=M_{\rm pk}/M_{\rm tr}$, and
power-law density profile exponent
(from top to bottom), $b=2.85$
(first block); $b=2.90$ (second block);
$b=2.95$ (third block); $b=3$ (fourth
block).}
\label{t:kab}
\end{table}
The dependence on the power-law density profile
exponent, $b$, for fixed ratio of homogeneous
peak mass to total mass within the truncation
radius, $\kappa$, is steeper for low $\kappa$
and vice versa.

For proto-galaxies $(M/{\rm M}_{10}=100)$,
slopes $b<2.85$
would imply exceedingly large initial peak
overdensities $(\delta_{\rm pk}>0.74)$,
and slopes $b>2.996$ would imply exceedingly large
cosmic times at the end of central collapse
$(t_c>0.9\,{\rm Gyr})$.   Mass ratios, $\kappa$,
outside the above mentioned range
would be inconsistent with Eq.\,(\ref{eq:Mhd}),
related to the cases of interest deduced from
observations and/or supported by theoretical
arguments.   The homogeneous density profile,
$b=3$, makes a useful limiting case.

In the current model, the hole mass coincides
with the homogeneous peak mass, and $\kappa$
can be deduced from Eq.\,(\ref{eq:Mhd}).
As a reference case, the values $(\beta,z)=
(0.65, 6/29)$ shall be chosen, which make
Eq.\,(\ref{eq:Mhd}) coincide with its
empirical counterpart deduced from a sample
of 37 spiral galaxies (Ferrarese, 2002),
together with the value,
$b=2.90$ in Eq.\,(\ref{eq:ropi}).
The related values of the homogeneous peak
radius, $r_{\rm pk}$, and the ratio of the
homogeneous peak radius to truncation radius,
$\Xi^{-1}=r_{\rm pk}/R_{\rm tr}$, are listed
in Tab.\,\ref{t:MR} for different masses at
the beginning of evolution, assumed to be at
recombination epoch.
\begin{table}
\begin{tabular}{clllll}
\hline
\hline
\multicolumn{1}{c}{$\log(M/{\rm M}_{10})$} &
\multicolumn{1}{c}{$R$/kpc} &
\multicolumn{1}{c}{$r_{\rm pk}$/kpc}
&
\multicolumn{1}{c}{$\overline{\delta}_{\rm M}$} &
\multicolumn{1}{c}{$\kappa$} &
\multicolumn{1}{c}{$\Xi^{-1}$} \\
\hline
          $-$1 & 1.72 E$-$1 & 6.98 E$-$4 & 1.27 E$-$2 & 1.12 E$-$7 & 4.06 E$-$3 \\
$\phantom{-}$0 & 3.70 E$-$1 & 2.52 E$-$3 & 1.00 E$-$2 & 5.01 E$-$7 & 6.80 E$-$3 \\
$\phantom{-}$1 & 7.98 E$-$1 & 9.10 E$-$3 & 7.26 E$-$3 & 2.24 E$-$6 & 1.14 E$-$2 \\
$\phantom{-}$2 & 1.72 E$+$0 & 3.28 E$-$2 & 4.81 E$-$3 & 1.00 E$-$5 & 1.91 E$-$2 \\
$\phantom{-}$3 & 3.70 E$+$0 & 1.18 E$-$1 & 3.08 E$-$3 & 4.47 E$-$5 & 3.20 E$-$2 \\
$\phantom{-}$4 & 7.98 E$+$0 & 4.28 E$-$1 & 1.72 E$-$3 & 1.99 E$-$4 & 5.36 E$-$2 \\
$\phantom{-}$5 & 1.72 E$+$1 & 1.54 E$+$0 & 7.89 E$-$4 & 8.91 E$-$4 & 8.98 E$-$2 \\
$\phantom{-}$6 & 3.70 E$+$1 & 5.57 E$+$0 & 2.90 E$-$4 & 3.98 E$-$3 & 1.50 E$-$1 \\

\hline\hline
\end{tabular}
\caption{The homogeneous peak radius,
$r_{\rm pk}$, the ratio of homogeneous
peak radius
to truncation radius, $\Xi^{-1}=r_{\rm pk}
/R_{\rm tr}$, and the ratio of homogeneous
peak mass to total mass within the truncation
radius, $\kappa=M_{\rm pk}/M_{\rm tr}$, for
overdensities with different masses, at
recombination epoch, density profile defined
by Eq.\,(\ref{eq:ropi}), $\kappa$ deduced
from Eq.\,(\ref{eq:Mhd}), particularized
to the selected reference case, $b=2.90$
and $(\beta,z)=(0.65, 6/29)$.   The
rms overdensity value, $(\overline{\delta}_
M)_i=\overline{\delta}_M$, and the overdensity
radius, $R_i=R$, are taken from Tab.\,\ref{t:CDM} for
sake of completeness.}
\label{t:MR}
\end{table}
Initial radii, $R_i=R$, and rms overdensity
values, $(\overline{\delta}_{\rm M})_i=
\overline{\delta}_{\rm M}$, are taken
from Tab.\,\ref{t:CDM} for sake of
completeness.   The above mentioned
quantities are independent of the overdensity
height.

The homogeneous peak overdensity,
$\delta_{\rm pk}$, the local
overdensity at the truncation radius,
$\delta_{\rm tr}$, the homogeneous
peak density, $\rho_{\rm pk}$, the
overdensity turnaround radius, $R_
{\rm max}$, the cosmic time at
turnaround, $t_{\rm max}$, and at
the end of central collapse, $t_c$,
are listed in Tab.\,\ref{t:evo} for
mean overdensity heights, $\overline{\nu}_
i=1,2,3,4$ (from top to bottom in
each bloch), and different masses,
with regard to the selected reference
case.
\begin{table}
\begin{tabular}{lllllll}
\hline
\hline
\multicolumn{1}{c}{$\log\frac M{{\rm M}_{10}}$} &
\multicolumn{1}{c}{$\delta_{\rm pk}$} &
\multicolumn{1}{c}{$\delta_{\rm tr}$}
&
\multicolumn{1}{c}{$\frac{\rho_{\rm pk}}{({\rm M}_{10}/{\rm kpc}^3)}$} &
\multicolumn{1}{c}{$\frac{r_{\rm max}}{\rm kpc}$} &
\multicolumn{1}{c}{$\frac{t_{\rm max}}{\rm Gyr}$} &
\multicolumn{1}{c}{$\frac{t_c}{\rm Gyr}$} \\
\hline
$-$1 & 6.98 E$-$1 & $-$2.11 E$-$2 & 7.97 E$+$0 & 1.37 E$+$1 & 8.51 E$-$1 & 1.55 E$-$2 \\
     & 7.19 E$-$1 & $-$8.78 E$-$3 & 8.07 E$+$0 & 6.54 E$+$0 & 3.05 E$-$1 & 1.53 E$-$2 \\
     & 7.40 E$-$1 & $+$3.49 E$-$3 & 8.17 E$+$0 & 4.69 E$+$0 & 1.68 E$-$1 & 1.51 E$-$2 \\
     & 7.62 E$-$1 & $+$1.58 E$-$2 & 8.27 E$+$0 & 3.56 E$+$0 & 1.10 E$-$1 & 1.49 E$-$2 \\
     &            &               &            &            &           &            \\
$\phantom{-}$0 & 6.08 E$-$1 & $-$2.37 E$-$2 & 7.55 E$+$0 & 3.75 E$+$1 & 1.22 E$+$0 & 1.66 E$-$2 \\
     & 6.24 E$-$1 & $-$1.40 E$-$2 & 7.62 E$+$0 & 1.89 E$+$1 & 4.35 E$-$1 & 1.64 E$-$2 \\
     & 6.40 E$-$1 & $-$4.40 E$-$3 & 7.70 E$+$0 & 1.27 E$+$1 & 2.39 E$-$1 & 1.62 E$-$2 \\
     & 6.56 E$-$1 & $+$5.24 E$-$3 & 7.77 E$+$0 & 9.65 E$+$0 & 1.57 E$-$1 & 1.60 E$-$2 \\
     &            &               &            &            &           &            \\
$\phantom{-}$1 & 5.23 E$-$1 & $-$2.63 E$-$2 & 7.15 E$+$0 & 1.11 E$+$2 & 1.96 E$+$0 & 1.82 E$-$2 \\
     & 5.34 E$-$1 & $-$1.93 E$-$2 & 7.20 E$+$0 & 5.58 E$+$1 & 6.98 E$-$1 & 1.80 E$-$2 \\
     & 5.45 E$-$1 & $-$1.23 E$-$2 & 7.25 E$+$0 & 3.75 E$+$1 & 3.83 E$-$1 & 1.77 E$-$2 \\
     & 5.56 E$-$1 & $-$5.28 E$-$3 & 7.31 E$+$0 & 2.83 E$+$1 & 2.50 E$-$1 & 1.75 E$-$2 \\
     &            &               &            &            &           &            \\
$\phantom{-}$2 & 4.43 E$-$1 & $-$2.87 E$-$2 & 6.77 E$+$0 & 3.59 E$+$2 & 3.63 E$+$0 & 2.04 E$-$2 \\
     & 4.50 E$-$1 & $-$2.40 E$-$2 & 6.81 E$+$0 & 1.81 E$+$2 & 1.29 E$+$0 & 2.02 E$-$2 \\
     & 4.57 E$-$1 & $-$1.93 E$-$2 & 6.84 E$+$0 & 1.21 E$+$2 & 7.05 E$-$1 & 2.00 E$-$2 \\
     & 4.64 E$-$1 & $-$1.47 E$-$2 & 6.87 E$+$0 & 9.11 E$+$1 & 4.60 E$-$1 & 1.98 E$-$2 \\
     &            &               &            &            &           &            \\
$\phantom{-}$3 & 3.68 E$-$1 & $-$3.03 E$-$2 & 6.42 E$+$0 & 1.20 E$+$3 & 7.05 E$+$0 & 2.36 E$-$2 \\
     & 3.72 E$-$1 & $-$2.74 E$-$2 & 6.44 E$+$0 & 6.04 E$+$2 & 2.50 E$+$0 & 2.34 E$-$2 \\
     & 3.76 E$-$1 & $-$2.44 E$-$2 & 6.46 E$+$0 & 4.04 E$+$2 & 1.36 E$+$0 & 2.32 E$-$2 \\
     & 3.81 E$-$1 & $-$2.14 E$-$2 & 6.48 E$+$0 & 3.04 E$+$2 & 8.89 E$-$1 & 2.29 E$-$2 \\
     &            &               &            &            &           &            \\
$\phantom{-}$4 & 2.97 E$-$1 & $-$3.17 E$-$2 & 6.09 E$+$0 & 4.64 E$+$3 & 1.68 E$+$1 & 2.84 E$-$2 \\
     & 3.00 E$-$1 & $-$3.00 E$-$2 & 6.10 E$+$0 & 2.32 E$+$3 & 5.97 E$+$0 & 2.82 E$-$2 \\
     & 3.02 E$-$1 & $-$2.83 E$-$2 & 6.11 E$+$0 & 1.55 E$+$3 & 3.25 E$+$0 & 2.80 E$-$2 \\
     & 3.04 E$-$1 & $-$2.67 E$-$2 & 6.12 E$+$0 & 1.17 E$+$3 & 2.12 E$+$0 & 2.79 E$-$2 \\
     &            &               &            &            &           &            \\
$\phantom{-}$5 & 2.31 E$-$1 & $-$3.25 E$-$2 & 5.78 E$+$0 & 2.18 E$+$4 & 5.43 E$+$1 & 3.64 E$-$2 \\
     & 2.32 E$-$1 & $-$3.18 E$-$2 & 5.78 E$+$0 & 1.09 E$+$4 & 1.92 E$+$1 & 3.63 E$-$2 \\
     & 2.33 E$-$1 & $-$3.10 E$-$2 & 5.79 E$+$0 & 7.28 E$+$3 & 1.05 E$+$1 & 3.61 E$-$2 \\
     & 2.34 E$-$1 & $-$3.02 E$-$2 & 5.79 E$+$0 & 5.46 E$+$3 & 6.81 E$+$0 & 3.60 E$-$2 \\
     &            &               &            &            &           &            \\
$\phantom{-}$6 & 1.69 E$-$1 & $-$3.29 E$-$2 & 5.49 E$+$0 & 1.28 E$+$5 & 2.43 E$+$2 & 5.13 E$-$2 \\
     & 1.69 E$-$1 & $-$3.26 E$-$2 & 5.49 E$+$0 & 6.39 E$+$4 & 8.61 E$+$1 & 5.11 E$-$2 \\
     & 1.69 E$-$1 & $-$3.24 E$-$2 & 5.49 E$+$0 & 4.26 E$+$4 & 4.69 E$+$1 & 5.10 E$-$2 \\
     & 1.70 E$-$1 & $-$3.21 E$-$2 & 5.49 E$+$0 & 3.19 E$+$4 & 3.04 E$+$1 & 5.09 E$-$2 \\

\hline\hline
\end{tabular}
\caption{Homogeneous peak overdensity,
$\delta_{\rm pk}$, local overdensity
at truncation radius, $\delta_{\rm tr}$,
homogeneous peak density, $\rho_{\rm pk}$,
overdensity turnaround radius, $R_{\rm max}$,
cosmic time at turnaround, $t_{\rm max}$, and
at the end of
central collapse, $t_c$, for mean
overdensity heights, $\overline{\nu}_i=1,2,3,4$
(from top to bottom of each block), and
different masses.   The values are related
to the selected reference case, as in
Tab.\,\ref{t:MR}.}
\label{t:evo}
\end{table}
The homogeneous peak density,
$\rho_{\rm pk}$, and the local
overdensity at truncation radius,
$\delta_{\rm tr}$, decrease only
slowly with increasing masses,
and the same holds for the homogeneous
peak overdensity, $\delta_{\rm pk}$,
though to a slightly larger extent.
In particular, the model density
profile, expressed by Eq.\,(\ref
{eq:ropi}), implies the existence
of an underdense external region
(negative overdensity) up to the
truncation radius, at the end of
central collapse.   Both the
overdensity turnaround radius,
$r_{\rm max}$, and the cosmic time
at the end of central collapse,
$t_c$, decrease for increasing
mean overdensity height, $\overline{\nu}_
i$.   Accordingly, in the light
of the model under discussion, hole
formation occurs within a few
hundredths of Gyr, or $z>10$.

\begin{figure}
\centerline{\psfig{file=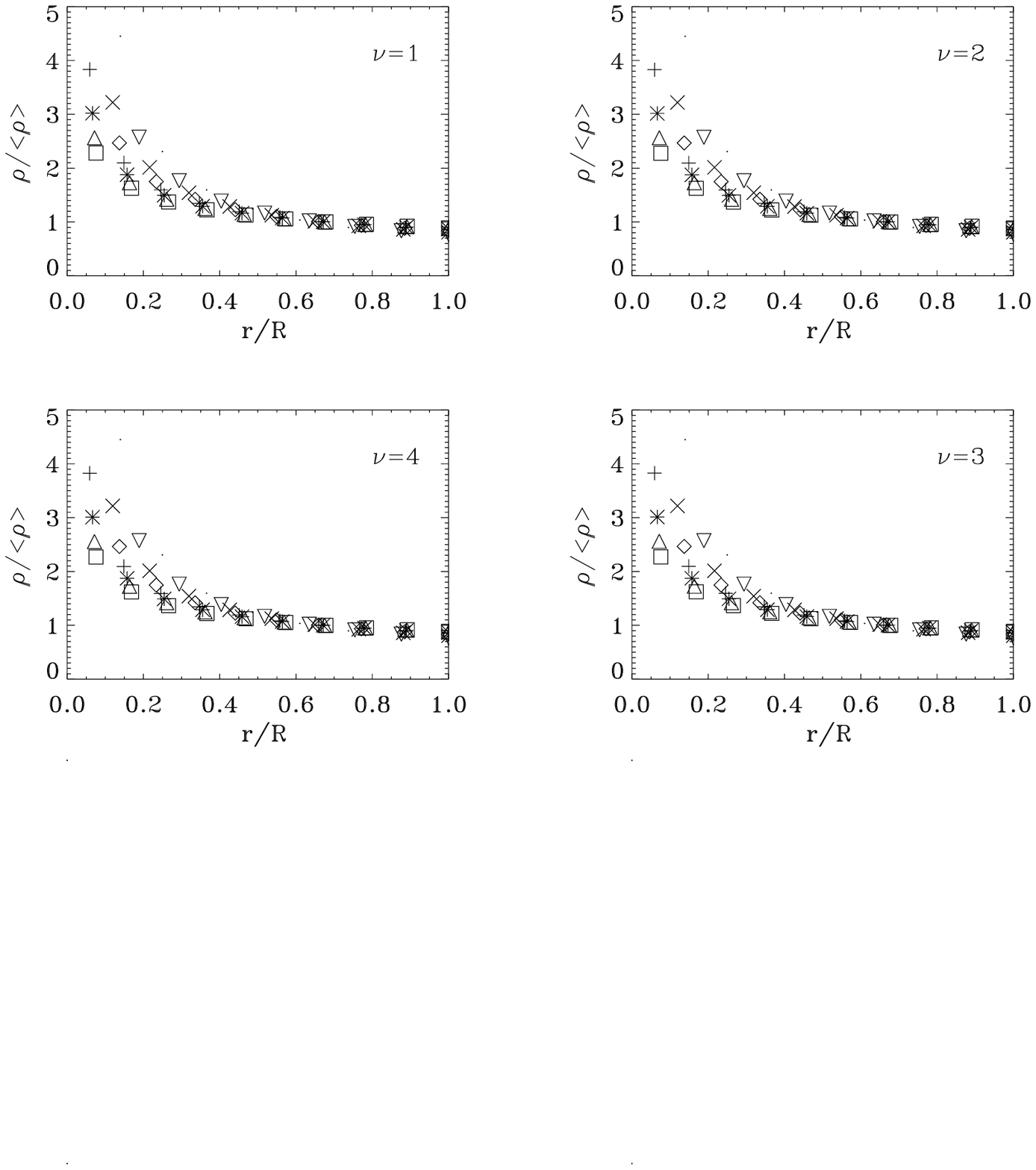,height=100mm,width=90mm}}
\caption{The fractional density,
$\rho(r)/\overline{\rho}=\rho(r)/<\rho>$,
as a function of the fractional radius, $\xi=r/R$,
at the end of central collapse, for different
masses and mean overdensity heights, $\overline
{\nu}_i=\nu$, with regard to the selected
reference case.   Different symbols correspond
to the following values of $\log(M/{\rm M}_{10})$:
$-$1 (squares); 0 (triangles); 1 (asterisks);
2 (Greek crosses); 3 (diamonds); 4 (St. Andrew's
crosses); 5 (reversed triangles); 6 (dots).}
\label{f:fdfr}    
\end{figure}

The fractional density at the end
of central collapse, $\rho/\overline
{\rho}$, as a function of the
fractional radius, $\xi=r/R$,
at the end of central collapse,
for different masses and mean
overdensity heights, with regard to
the selected reference case,
is plotted in Fig.\,\ref{f:fdfr}.
The density profile of the outer shells,
which are still expanding together with
the universe, exhibits no dependence on
the overdensity mass.   The contrary holds for
the inner shells where, for assigned
fractional radius, the fractional density
is an increasing function of the overdensity
mass.   The dependence on the mean overdensity
height appears to be negligible.

\begin{figure}
\centerline{\psfig{file=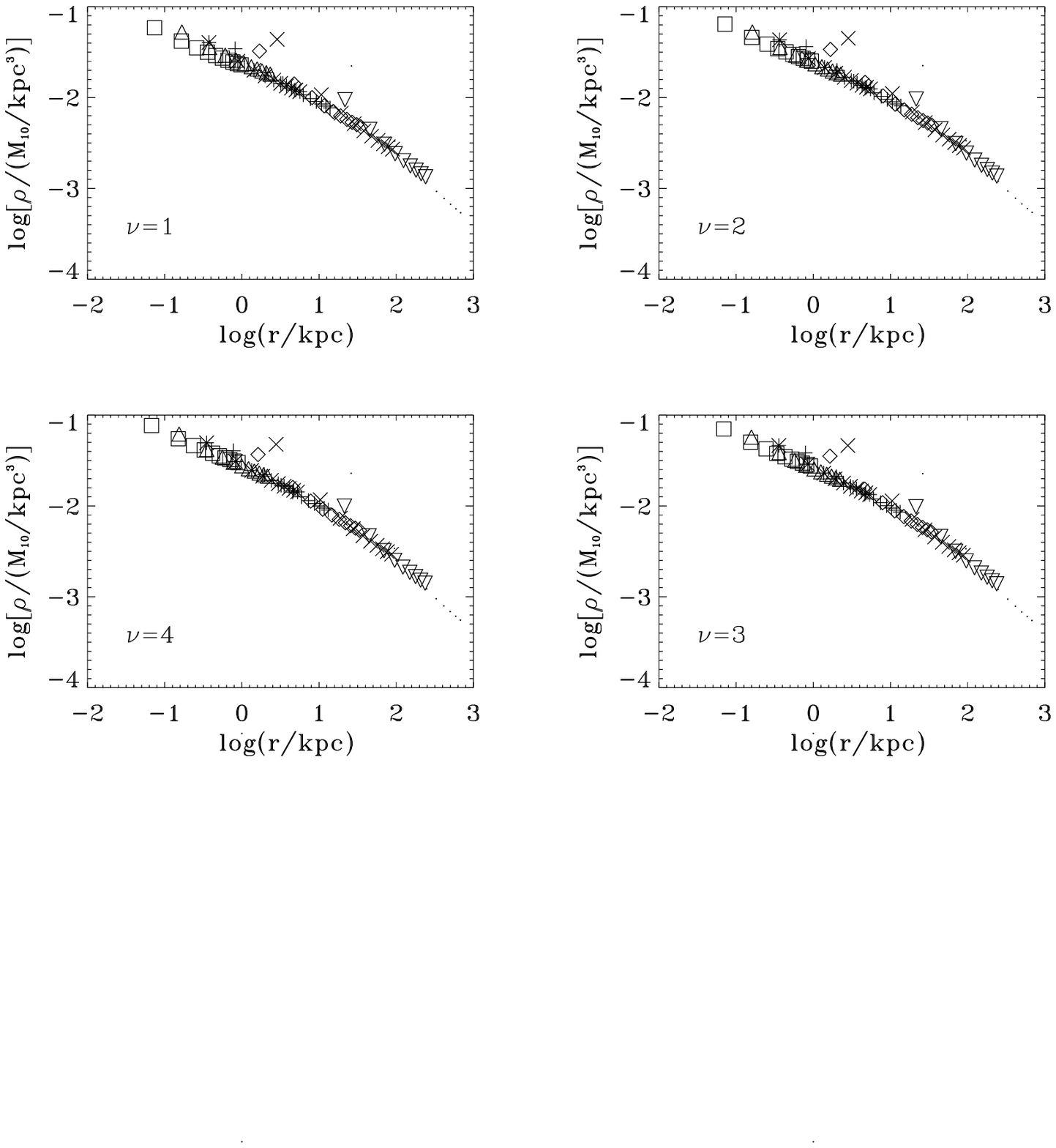,height=100mm,width=90mm}}
\caption{The decimal logarithm of the density,
$\log[\rho/({\rm M}_{10}/{\rm kpc}^3)]$, as a
function of the decimal logarithm of the
radius, $\log(r/{\rm kpc})$,
at the end of central collapse, for different
masses and mean overdensity heights, $\overline
{\nu}_i=\nu$,
with regard to the selected reference case.
Caption of symbols as in
Fig.\,\ref{f:fdfr}.}
\label{f:ldr}
\end{figure}

The density, $\rho$, as a function of the
radius, $r$, at the end of central collapse,
for different masses and mean overdensity heights,
with regard to the selected reference case,
is plotted on a logarithmic plane in
Fig.\,\ref{f:ldr}.
The density profile is universal, with
the exception of the inner shells, where
a higher density is attained.
The dependence on the mean overdensity
height appears to be negligible.

\begin{figure}
\centerline{\psfig{file=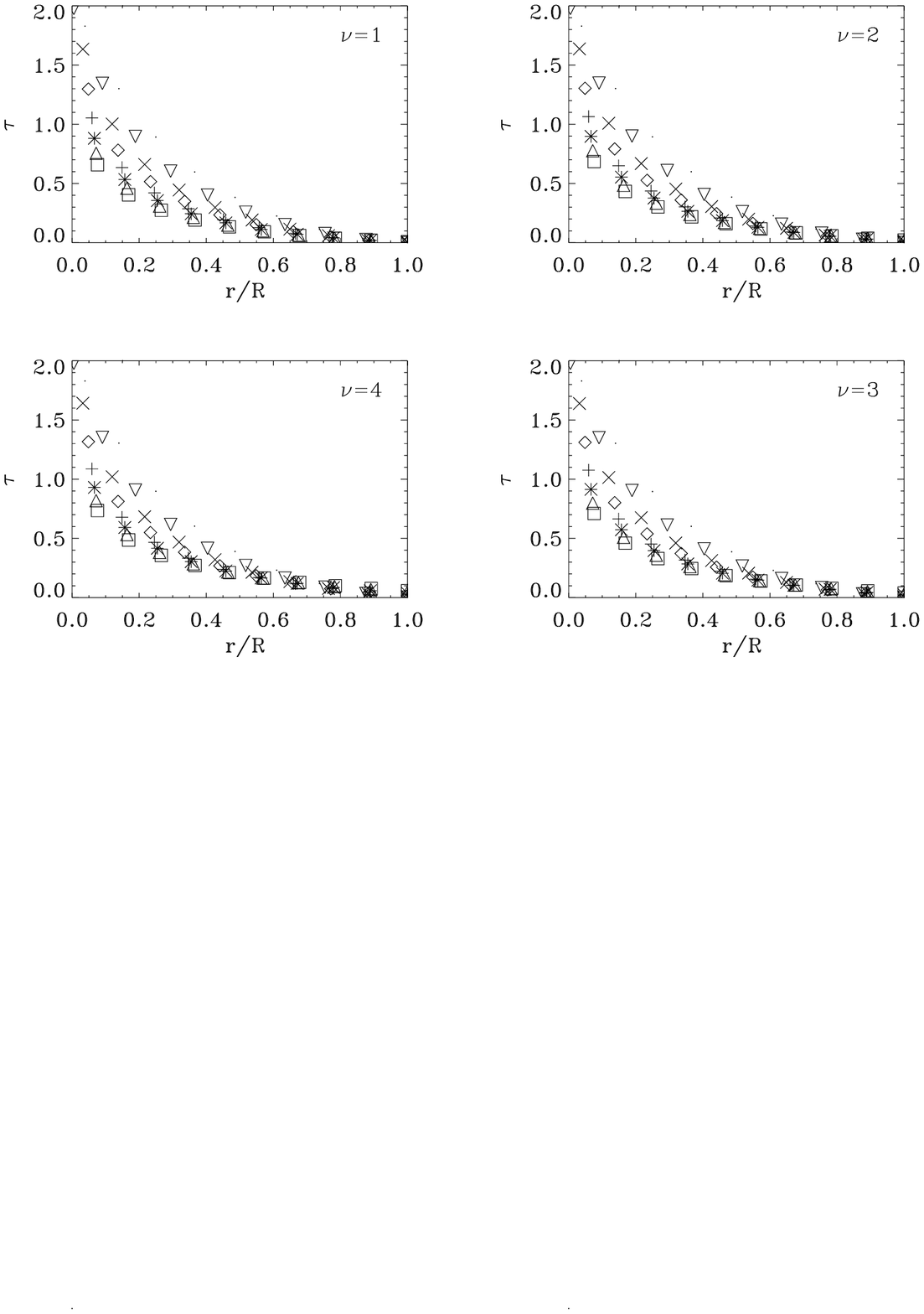,height=100mm,width=90mm}}
\caption{The dimensionless time, $\tau=\tau_c(r_i)=
t_c/t_{\rm max}(r_i)$, as a function of the
fractional radius, $\xi=r/R$, at the end
of central collapse,
for different masses and mean overdensity heights,
with regard to the selected reference case.
Caption of symbols as in Fig.\,\ref{f:fdfr}.}
\label{f:tauc}
\end{figure}

The dimensionless time, $\tau=\tau_c(r_i)=
t_c/t_{\rm max}(r_i)$, as a function of the
fractional radius, $\xi=r/R$, at the end
of central collapse,
for different masses and mean overdensity heights,
with regard to the selected reference case,
is plotted in Fig.\,\ref{f:tauc}.
The dimensionless time related to the
outer shells, which are still expanding
together with the universe $(\tau\appgeq
0)$, exhibits no dependence on overdensity
mass.   The contrary holds for the inner
shells $(\tau\appgeq0.5)$ where, for
assigned fractional radius, the dimensionless
time is an increasing function of the
overdensity mass.   In other words,
with respect to a selected fractional radius,
the homogeneous peak collapses ``faster''
in low-mass overdentity than in large-mass
overdensities.
The dependence on the mean overdensity
height appears to be negligible.

\begin{figure}
\centerline{\psfig{file=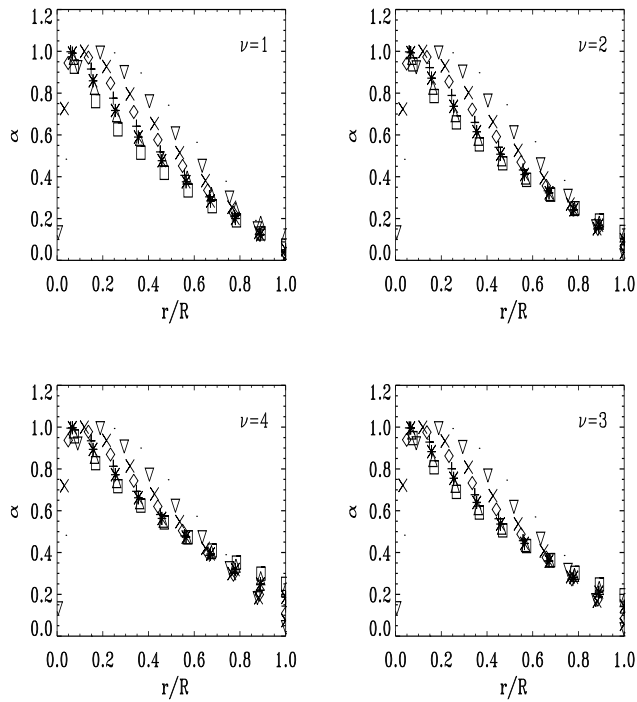,height=100mm,width=90mm}}
\caption{The dimensionless distance, $\alpha=
\alpha_c(r_i)=
r(t_c)/r_{\rm max}(r_i)$, as a function of the
fractional radius, $\xi=r/R$, at the end
of central collapse,
for different masses and mean overdensity heights,
with regard to the selected reference case.
Caption of symbols as in Fig.\,\ref{f:fdfr}.}
\label{f:alfc}
\end{figure}

The dimensionless distance, $\alpha=\alpha
_c(r_i)=r(t_c)/r_{\rm max}(r_i)$, as a function
of the fractional radius, $\xi=r/R$, at the end
of central collapse,
for different masses and mean overdensity heights,
with regard to the selected reference case,
is plotted in Fig.\,\ref{f:alfc}.
The dimensionless distance related to the
outer shells, which are still expanding
together with the universe $(\alpha\appgeq
0)$, is a decreasing function of the overdensity
mass, according to the initial conditions
via Eq.\,(\ref{eq:alc}).   The above mentioned
effect is less evident moving inwards, until
no appreciable mass dependence is shown.
The trend is reversed for the still expanding
inner shells where the dimensionless distance
is an increasing function of the overdensity
mass.   For fixed fractional radius, $r(t_c)/
R(t_c)$, the inner shells are closer to
maximum expansion, or end of contraction,
for large-mass overdensities with respect
to low-mass overdensities.   In other words,
for shells with fixed fractional radius,
the homogeneous peak collapses ``faster''
in low-mass overdensities with respect to
large-mass overdensities: at the end of
central collapse, shells which are turning
around correspond to a smaller initial
fractional radius for low-mass overdensities,
with respect to large-mass overdensities.
The fractional radius at the end of central
collapse, $\xi=r/R$, where the fractional
distance, $\alpha$, shows no appreciable
dependence on the overdensity mass, is a
decreasing function of the mean overdensity height,
passing from $\xi\approx0.9$ for $\overline
{\nu}_i=1$ to $\xi\approx0.7$ for $\overline
{\nu}_i=4$.

\begin{figure}
\centerline{\psfig{file=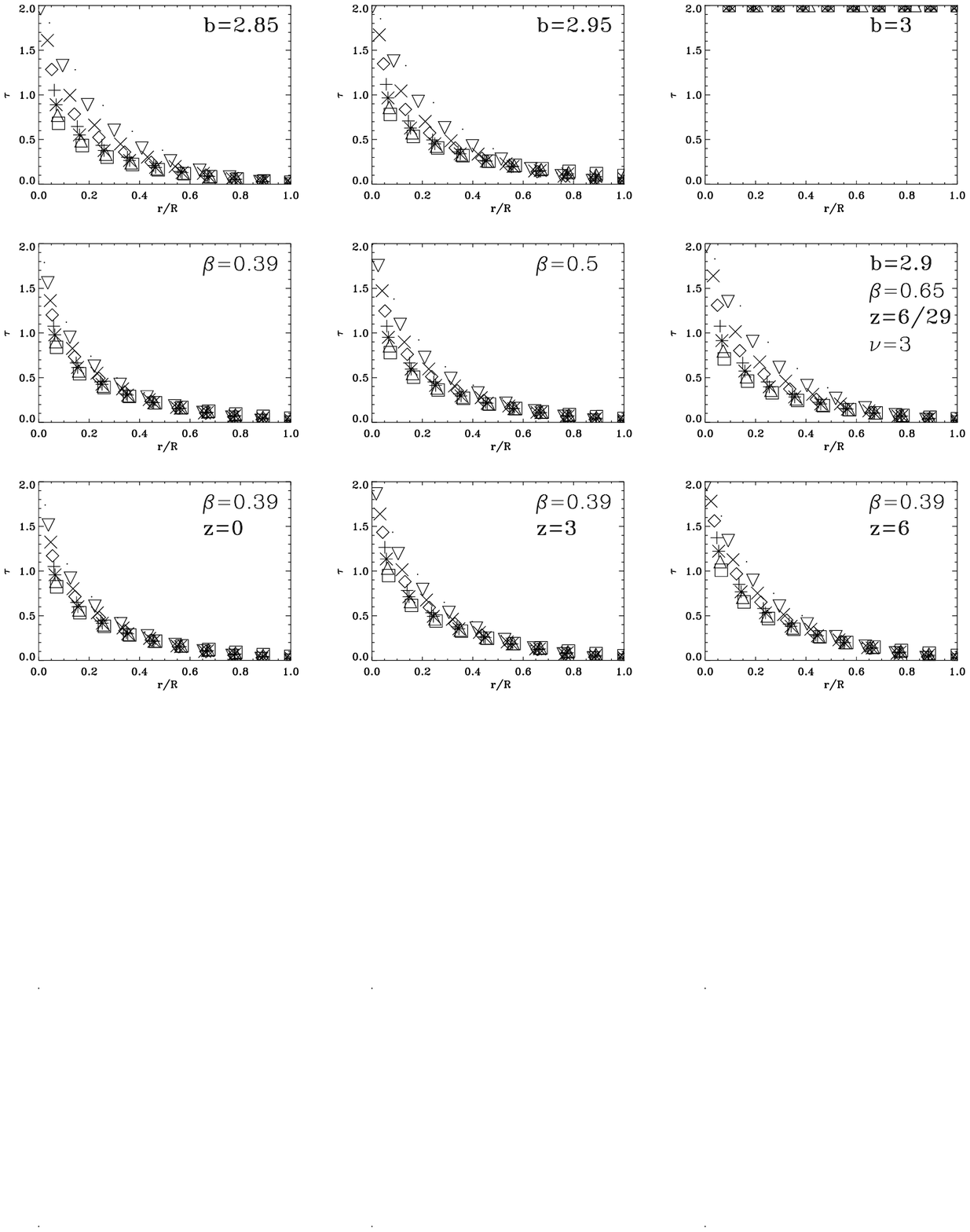,height=100mm,width=90mm}}
\caption{The dimensionless time, $\tau=\tau_c(r_i)=
t_c/t_{\rm max}(r_i)$, as a function of the
fractional radius, $\xi=r/R$, at the end
of central collapse,
for different masses, mean overdensity height $\nu=
\overline{\nu}=3$, and different slopes of
the initial density profile, $b$, Eq.\,(\ref
{eq:ropi}), exponent, $\beta$, and redshift,
$z$, appearing in the assumed hole-hosting
dark matter halo mass relation, Eq.\,(\ref
{eq:Mhd}).   The selected reference
case is placed on the middle right (same as in
Fig.\,\ref{f:tauc}, bottom right).   The
parameter values which have been changed
with respect to the reference case, are
listed on each panel.   The limiting situation
of homogeneous initial density profiles, $b=3$,
is shown for comparison only.
Caption of symbols as in Fig.\,\ref{f:fdfr}.}
\label{f:tau3}
\end{figure}

\begin{figure}
\centerline{\psfig{file=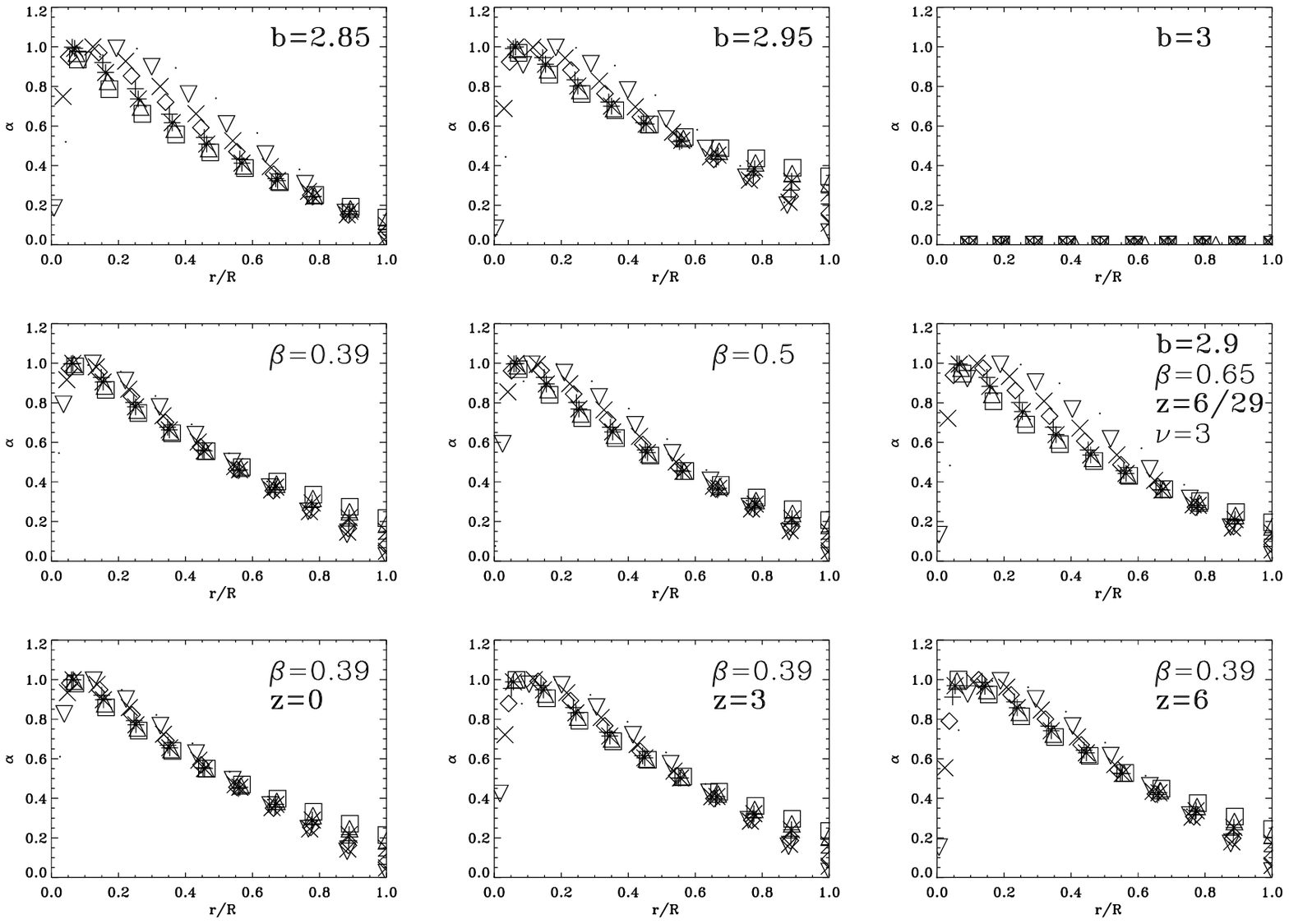,height=100mm,width=90mm}}
\caption{The dimensionless distance, $\alpha=
\alpha_c(r_i)=
r(t_c)/r_{\rm max}(r_i)$, as a function of the
fractional radius, $\xi=r/R$, at the end
of central collapse,
for different masses, mean overdensity height $\nu=
\overline{\nu}=3$, and different slopes of
the initial density profile, $b$, Eq.\,(\ref
{eq:ropi}), exponent, $\beta$, and redshift,
$z$, appearing in the assumed hole-hosting
dark matter halo mass relation, Eq.\,(\ref
{eq:Mhd}).   The selected reference
case is placed on the middle right (same as in
Fig.\,\ref{f:alfc}, bottom right).   The
parameter values which have been changed
with respect to the reference case, are
listed on each panel.   The limiting situation
of homogeneous initial density profiles, $b=3$,
is shown for comparison only.
Caption of symbols as in Fig.\,\ref{f:fdfr}.}
\label{f:alf3}
\end{figure}

The effect of a change of the model
parameters on the results, is shown
in Fig.\,\ref{f:tau3} for the
dimensionless time, $\tau=\tau_c(r_i)=
t_c/t_{\rm max}(r_i)$, as a function of the
fractional radius, $\xi=r/R$, at the end
of central collapse, and in Fig.\,\ref{f:alf3}
for the dimensionless distance, $\alpha=
\alpha_c(r_i)=
r(t_c)/r_{\rm max}(r_i)$, as a function of the
fractional radius, $\xi=r/R$, at the end
of central collapse, respectively,
for different masses.
The selected reference case is placed on
the middle right, which is the same as in
Fig.\,\ref{f:tauc} and Fig.\,\ref{f:alfc}
(bottom right), respectively.   The changes
in parameter values with respect to the
selected reference case, are listed on
each panel.

The limiting situation of
homogeneous initial density profiles,
$b=3$, is shown for comparison only, as
it would imply a cosmic time at the end
of central collapse, longer than the age
of the universe for large-mass overdensities
$(M/{\rm M}_{10}\ge10^3)$.   Slopes within
a fiducial range, $2.85\le b\le2.95$, make
little change on both the dimensionless
time and the dimensionless distance, as shown
in Figs.\,\ref{f:tau3} and \ref{f:alf3}.
A weak trend is exhibited, where the evolution
of different shells is closer in shallower
density profiles and vice versa, as expected.

According to the assumed
hole-hosting dark matter halo mass
relation, Eq.\,(\ref{eq:Mhd}), the hole mass
is larger (for assigned hosting dark matter
halo mass) for
larger exponent, $\beta$, and/or redshift,
$z$, and vice versa, which implies closer
evolution of different shells for low-mass
holes and vice versa.   An inspection of
Figs.\,\ref{f:tau3} and \ref{f:alf3} shows
that it is the case, and the related trend
is more visible than in the case of changing
slope of the initial density profile, $b$.

In the light of the model presented in the
current attempt, holes are the first structures
which form in proto-galaxies, when the upper
shells are still collapsing or expanding.

\section{Discussion}
\label{disc}

Gravitational collapse may be a viable
mechanism for hole generation, due to
two orders of reasons.   First, remnants
of sufficiently massive Type II supernovae
are thought to be black holes (e.g.,
Nomoto et al. 2005; Ohkube et al. 2006).
Second, the mean density of a
(spherical-symmetric) black hole is a
decreasing function of the gravitational
radius.   Low-mass $(m\appleq20{\rm m}_
\odot)$ Type II supernovae leave a
neutron (or more exotic) star remnant,
which is sustained by the Fermi pressure
within a radius about three times larger
than the gravitational radius.   The Fermi
pressure makes the collapsing core bounce
at a density about twice the nuclear density,
and later attain a virialized configuration.
On the other hand, the bounce would occur inside
the gravitational radius for more massive
$(m\approx20\,$-$25{\rm m}_\odot)$ Type II
supernovae, which implies black hole
formation.   If (nonbaryonic) dark matter
is mainly made of fermions, a similar
mechanism could hold.   Low-mass $(M\appleq
10{\rm M}_{10})$ dark matter overdensities
could leave a fermion ball remnant at the
end of central collapse, while larger
masses could imply hole generation (e.g.,
Viollier 1994; Munyaneza and Viollier
2002; Munyaneza and Biermann 2006).

Radial motions are better preserved in
collisional fluids where energy dissipation
takes place, such as in supernovae, than in
collisionless fluids where no energy
dissipation occurs, such as dark matter
overdensities.   Clump formation and
tidal interactions between clumps or
from neighbourhing overdensities, convert
ordered radial motions into random orbital
motions.   On the other hand, the local
maximum at the overdensity deep interior,
is a very special place where both tidal
interactions and acquisition of angular
momentum may safely be neglected.
Accordingly, radial motions could be
preserved up to hole formation.


The extrapolation of the hole-hosting dark
matter halo mass relation, Eq.\,(\ref{eq:Mhd}),
to early cosmic times, has been considered as
the most viable expression of the homogeneous
peak to overdensity
mass ratio.   The related value could be
overstimated if only the inner part of the
homogeneous peak is the hole progenitor,
and understimated if substantial mass accretion
onto the hole works.   The extrapolation of
Eq.\,(\ref{eq:Mhd}) in the above mentioned
sense, implies the mass of the hole progenitor
is proportional to the overdensity mass, at
fixed redshift.
If the total clump number or the typical
clump mass does not appreciably depend on
the overdensity mass, clumps within low-mass
overdensities are expected to be efficient
in exerting tidal actions and related departure
from radial motions.   Accordingly, central
collapse and hole formation could be inhibited,
in alternative to the formation of a fermion ball.

Though the assumed density profile at the
beginning of evolution, Eq.\,(\ref{eq:ropi}),
has no physical motivation, still it cannot
be considered as completely {\it ad hoc}.
In fact, it is lying between two limiting
situations, related to a homogeneous and
a generalized Roche sphere (a homogeneous
sphere surrounded by a massless atmosphere),
respectively.    It is defined by two
parameters: a power-law exponent, $b$,
and a fractional mass, $\kappa=M_{\rm pk}/
M_{\rm tr}$.   The special cases, $b=0$
and $b=3$, are related to generalized
Roche and homogeneous spheres, respectively.
A fiducial range, $2.85\le b\le2.95$, has
been chosen for the following reasons: low
values $(b<2.85)$ would imply exceedingly
high peak overdensities, while  large values
$(b>2.95)$ would imply exceedeingly long
cosmic times at the end of central collapse.
The main feature is the presence
of a homogeneous peak which turns around
and collapses into a black hole, and an
inhomogeneous envelope which, after
expansion, collapses and relaxes.

The fractional mass, $\kappa$, has been
selected as a function of both the
overdensity mass and the redshift,
expressed by Eq.\,(\ref{eq:Mhd}),
provided overdensity and
homogeneous peak mass coincide with
dark halo and hole mass, respectively.
Recent empirical correlations are
related to special choices of the
parameters in Eq.\,(\ref{eq:Mhd}),
namely: $(\beta,z)=(0.65,6/29)$ 
(Ferrarese, 2002); $(\beta,z)=
(0.27,19/58)$ (Baes et al., 2003);
$\beta=0.39$
(Shankar and Mathar, 2007).  Also,
Eq.\,(\ref{eq:Mhd}) is close to its
counterpart calibrated locally
through statistical arguments
(Shankar et al., 2006) and to its
counterpart obtained using
semianalytical models (Granato et
al., 2004; Lapi et al., 2006).
In conclusion, the assumed density
profile at the beginning of evolution,
Eq.\,(\ref{eq:ropi}), may be considered
as a phenomenological one.

Strictly speaking, the model discussed
in the current attempt holds for isolated
overdensities in a secondary infall
scenario, where the inner and denser
regions first virialize while the
outer and less dense regions are still
expanding (Gunn 1977).   On the other
hand, the formation of dark haloes is
characterized by accretion of smaller
subunits or merger between systems of
comparable size, and identification
of dark matter haloes in simulations
appears problematic (e.g., Bett et al.
2007).   Unexpectedly, a recent
investigation shows that the secondary
infall scenario provides a valid
theoretical framework for calculating
the structure and evolution of dark
matter haloes in an expanding universe,
and that its predictions with respect
to the density profiles are in close
agreement with full $N$-body simulations
(Ascasibar et al. 2007).   In this
view, most of the diversity in
density profiles is contributed
by the scatter in their primordial
counterparts rather the scatter in angular
momentum (Ascasibar et al. 2007) i.e.
the nature of the
initial conditions rather than
the details of the evolution.

Though hole gravitational radius cannot
be resolved by numerical simulations,
the resulting dark halo density profile
appears to be cusped at the centre (e.g.,
Navarro et al. 1995, 1996; Moore et al.
1998, 1999; Fukushige and Makino 2001,
2003; Diemand et al. 2004; Reed et al.
2005), but recent results also allow
cored density profiles (e.g., Fukushige
and Makino 2004; Navarro et al. 2004;
Merritt et al. 2005).   If the presence
of a central cusp in simulated dark
matter haloes is a real effect instead
of an artefact due to computer codes,
a physical interpretation could be
hole formation at the end of central
collapse.

Central collapse within local density
maxima is one way of avoiding the
problem of haloes at high redshifts
$(z\approx6)$ that have had insufficient
time to grow at the Salpeter rate from
solar or intermediate mass black hole
progenitors (e.g., Willott et al., 2003;
Miller et al., 2006).
In this view, holes are the first
structures formed in evolving
overdensities, while proto-haloes,
proto-bulges, and proto-disks
are still relaxing.
The active (quasar) phase begins just
after hole formation, when the
outstanding shells are still collapsing,
and may occasionally be enhanced by
merger events.   The cosmic time at the
end of central collapse does not exceed
a few hundredths of Gyr, see Tab.\,\ref
{f:alfc}, in agreement with observations
of high-redshift quasars.   Virialization
takes place later, at a cosmic time equal
to about 3 Gyr, or $z\approx2$, provided
$M/{\rm M}_{10}=500$ [deduced from Cromm
et al., 2005, using $h=0.65$ in accordance
with Eq.\,(\ref{eq:copa})] and $\overline
{\nu}=3$ are typical for luminous quasars,
see again Tab.\,\ref
{f:alfc}.   The
occurrence of virialization, gas
exhaustion due to star formation,
and absence of merger events in
the hosting spheroid component,
mark the beginning of the quiescent
phase.

The simple model used in the
current attempt deals with
isolated overdensities where
neither accretion nor merging
occurs.   Hole formation via
dark matter collapse implies a
limited range of mass, where
the lower limit is related to
the stability of a fermion ball
(provided dark matter is mainly
made of fermions) or to the
occurrence of tidal effects
between clumps, and the upper
limit is related to the end of
central collapse at a cosmic
time equal to the age of the
universe.   Accordingly, very
massive holes are expected
within clusters of galaxies
and superclusters.   Though
the model is limited to hole
formation, still some qualitative
considerations on the evolution
can be performed.

The picture of holes forming at
high redshifts and remaining
largely unchanged since then, is
not consistent with the observed
luminosity density of active galactic
nuclei and the inferred local hole
density, unless the radiative efficiency
of the luminosity of active galactic nuclei
is unfeasibly high (Marconi et al., 2004).
So it appears that holes have been
continuing to accrete mass during
the cosmic epochs in which dark matter
haloes and their associated galaxies
have also continued to be assembled.

In the light of the model used in the
current paper, the above scenario
still holds with regard to the baryonic
precursor of the hole mass budget.   On
the other hand, little change occurs
since hole formation with regard to the
(dominant) nonbaryonic precursor.
In fact, the basic
assumption is that seed holes are
created at the end of central collapse
$(z>10)$ instead of in major
mergers which, in addition, trigger
an episode of gas accretion in overdensities
with a pre-existing hole.   In other words,
mass accretion occurs
{\it after} hole formation, while in
earlier attempts mass accretion occurs
{\it during} hole formation (e.g., Di
Matteo et al., 2003; Bromley et al., 2004;
Miller et al., 2006; Lapi et al., 2006;
Li et al., 2007).
%
%
Baryon accretion may highly be reduced by
central starburst or radiative feedback
(e.g., Ciotti and Ostriker, 2007).
Nonbaryonic matter accretion may not
significantly increase hole mass, due
to the lack of a mechanism to dissipate
angular momentum, and it may be important
only in early times, $z\appgeq30$ (Mack
et al., 2007).

Several previous models have considered
the possibility that holes form from low-mass
black holes related to pop. II (e.g.,
Haiman and Loeb, 2001) or pop. III (e.g.,
Volonteri et al., 2003) stars.   However,
there are various problems associated with
the above mentioned scenarios (e.g.,
Haehnelt, 2003; Pelupessy et al., 2007).
More specifically, some
mechanism is required to facilitate the
migration of seed black holes to the centre
of their hosting spheroid.

Another class
of models focused on the possibility that
holes form directly from the collapse of a
large gaseous cloud placed in the central
overdensity region (e.g., Rees, 1984;
Haehnalt and Rees, 1993; Silk and Rees,
1998; Bromley et al., 2004), where the
problem is the need to avoid fragmentation
and related departure from radial motions
during the collapse.   In particular, it is
assumed (Bromley et al., 2004) that the
collapse is also connected to overdensity
major mergers, in the sense that the same
tidally stripped gas which falls into the
centre of a post-merger overdensity and
can fuel quasar accretion, is also viewed
as the source of the initial hole formation.

The model used in the current paper follows
the same line of thought but is different
to many respects, namely: (i) hole formation
is (mainly) due to nonbaryonic matter instead of gas;
(ii) the initial driving mechanism is contraction of
local density maxima after turnaround instead
of major overdensity merger; (iii) quasar
accretion is initially fuelled by gas secondary
infall instead of tidally stripped gas during
overdensity major mergers.

As reported in an earlier attempt (Bromley
et al., 2004), any potential model of quasar
and active galactic nucleus formation finds
itself faced with three major unknowns.
First: where, how and with what mass do the
initial holes form?   Secondly: what events
trigger their subsequent fuelling and growth,
how much fuel do they supply and how efficiently
is it converted into radiative energy?   Thirdly:
how does feedback (from star formation or the
active galactic nucleus activity itself) regulate
and possibly even check their growth?

The current paper answers the first point
raised above: hole formation is the result
of central collapse in local density
maxima, mainly due to nonbaryonic dark
matter.   Subsequent fueling, growth, and
feedback, are due to baryonic matter, the
sole which can radiate and make quasars
and active galactic nuclei shining.   On
the contrary, the hole mass budget is
mainly determined by its nonbaryonic
precursor.   In this view, current models
on active galactic nuclei formation and
co-evolution with hosting galaxies, are
expected still to hold with regard to the
second and the third point raised above.

\section{Conclusion}\label{conc}

Unsustained matter distributions unescapely collapse unless
fragmentation and centrifugal or pressure
support take place.   Starting from the above
evidence, supermassive compact objects at the
centre of large-mass galaxies (defined as
``holes'') have been conceived as the end-product
of the gravitational collapse of local
density maxima (defined as ``central collapse'')
around which positive density perturbations
(overdensities) are located.   At the beginning
of evolution, assumed to occur at recombination
epoch, local density maxima have been idealized as
homogeneous peaks, while the surrounding envelopes
have been described by a power-law density profile,
$\rho(r)\propto r^{b-3}$, $0\le b\le3$, where
$b=0$ represents a massless atmosphere and
$b=3$ a homogeneous layer.   The dependence
of the density profile on a second parameter,
chosen to be the ratio between peak and total
(truncated) mass, $\kappa=M_{\rm pk}/M_{\rm 
tr}$, has been analysed.

Overdensity evolution has been discussed
in the context of quintessence cosmological
models, which should be useful in dealing
with the virialized phase.   Aiming to describe
the central collapse, further investigation has been
devoted to a special case where the quintessence
effect is equivalent to additional curvature
$(w=-1/3)$, and overdensities exhibit the
selected density profile at recombination epoch.
A redshift-dependent, power-law relation between hole and
(nonbaryonic) dark halo mass has been
used to express the dependence of the fractional
mass, $\kappa=M_{\rm pk}/M_{\rm tr}$, on the
overdensity mass,
$M=M_{\rm tr}$, where the homogeneous peak and
overdensity mass are related to the hole and
dark halo mass, respectively.

Computations have been
performed for a wide range of masses,
$-1\le\log(M/{\rm M}_{10})\le6$, and mean overdensity
heights, $1\le\overline{\nu}_i\le4$, up to the
end of central collapse, for the following quantities:
homogeneous peak overdensity, $\delta_{\rm pk}$;
turnaround radius, $r_{\rm max}$; cosmic time
at the end of central collapse, $t_c$; density
profile, $\rho[r(t_c)]$; logarithmic density
profile, $\log[\rho/({\rm M}_{10}/{\rm kpc}^3)]$;
ratio of cosmic time at the end of central
collapse to turnaround time of the related
shell, $\tau_c=t_c/t[r_{\rm max}(r_i)]$; ratio
of shell radius at the end of central collapse
to shell radius at turnaround, $\alpha_c=r(t_c)/
r_{\rm max}(r_i)$.    With regard to the last
two above mentioned quantities, additional
computations have been performed for different
slopes of the initial density profile, $b$,
and different parameters appearing in the
hole-hosting dark matter halo mass relation,
$\beta$ and $z$, Eq.\,(\ref{eq:Mhd}).

The central collapse has been found to end in 
early times, no longer than a few hundredths 
of Gyr, which implies hole formation when 
proto-haloes, proto-bulges, and proto-disks are
still relaxing.   No appreciable
change has been found in the evolution (up
to the end of central collapse) of
different mean overdensity heights related to equal
masses.   On the other hand, it has been recognized
that homogeneous peaks collapse (in dimensionless
coordinates) ``faster'' with respect to
surroundings envelopes, in low-mass overdensities
than in large-mass overdensities.   In conclusion,
it has been inferred that gravitational collapse of
homogeneous peaks within overdensities may be
a viable mechanism for hole generation.

\section*{Acknowledgements}
Thanks are due to S. Masiero for fruitful
discussions.  The author is indebted with
M. Yu. Khlopov for informations on his and
coworkers' quoted papers about black hole
generation via topological defects during
the inflation epoch.  The author is deeply
grateful to an anonymous referee for
critical comments which made substantial
improvement to an earlier version of the
paper.

\appendix
\section*{Appendix}

\section{On the light velocity in vacuum}
\label{a:c}

Strictly speaking, light travels
across the vacuum (intended in
philosophical sense, as absence
of everything) only in classical
(Newton) and special-relativistic
(Einstein) mechanics, where it is
conceived as
made of corpuscles.    If, on the
other hand, light is conceived as
waves (Huygens), then it propagates
across a homogeneous medium called
ether.   In general-relativistic
mechanics (Einstein) ether is replaced by
gravitational field of the universe
as a whole.   In quantum mechanichs
(Bohr, de Broglie, Heisemberg,
Schr\"odinger)
light can be considered as made
of either corpuscles or waves,
and the propagation takes place
across the quantum void, which
hosts some form of energy, and
then has not to be intended in
the above specified philosophical
sense.   Light velocity is maximum
in vacuum (or ether, or quantum void),
while it
decreases for increasing refraction
index when propagation occurs
within a (homogeneous) medium,
due to photon interaction with
medium constituents.

Light velocity has been deduced
from experiments on the Earth
or within the solar system at most.
According to current QCDM cosmologies,
the solar system is embedded within
(nonbaryonic) dark matter and dark
energy.   The current value of the
light velocity coincides with the
maximum attainable value only if
no interaction is assumed to occur
between (i) photons and dark matter,
and (ii) photons and dark energy.
If otherwise, light velocity could
be higher in intergalactic voids
and/or in early (cosmic) times,
where the effect of the dark energy
was negligible.
Even in pure matter universes,
light propagation takes place
across the quantum void, where some
form of energy is present.

In conclusion, the
term ``light velocity in vacuum''
has to be intended as ``light
velocity in absence of baryonic
matter'' under the assumption
that photons interact with neither
(nonbaryonic) dark matter, nor
dark energy.

\section{The [kpc M$_{10}$ Gyr] system of measure}
\label{a:syme}

In dealing with large-scale celestial objects,
such as galaxies or cluster of galaxies, it may
be convenient to use an appropriate system of
measure instead of the standard astrophysical
one, [cm g s].   To this aim, unit length, unit
mass, and unit time, shall be taken as:
\begin{lefteqnarray}
\label{eq:kpc}
&& 1\,{\rm kpc}=3.085\,677\,580\,666\,31~10^{21}~{\rm cm}~~; \\
\label{eq:M10}
&& {\rm M}_{10}=10^{10}{\rm m}_\odot=1.989\,1~10^{43}~{\rm g}~~; \\
\label{eq:Gyr} 
&& 1\,{\rm Gyr}=10^9~{\rm y}=3.153\,6~10^{16}~{\rm s}~~;
\end{lefteqnarray}
and the related ``mathematical'' year is
defined as $1\,{\rm y}=(365\times24\times
60\times60){\rm s}$, independent of the
Earth revolution around the Sun.

According to Eq.\,(\ref{eq:kpc}), the
astronomical unit is:
\begin{equation}
\label{eq:AU}
1\,{\rm AU}=1.495\,978\,706\,91~10^{13}~{\rm cm}=
4.848\,136\,812\,10~10^{-9}~{\rm kpc}~~;
\end{equation}
using Eqs.\,(\ref{eq:kpc}) and (\ref{eq:Gyr}),
the velocity unit reads:
\begin{leftsubeqnarray}
\slabel{eq:va}
&& 1\,{\rm kpc}\,{\rm Gyr}^{-1}=0.978\,461\,942\,118\,946\,6~
{\rm km}\,{\rm s}^{-1}~~;  \\
\slabel{eq:vb}
&& 1\,{\rm km}\,{\rm s}^{-1}=1.022\,012\,156\,992\,443~{\rm kpc}\,
{\rm Gyr}^{-1}~~;
\label{seq:v}
\end{leftsubeqnarray}
and the light velocity in vacuum is:
\begin{equation}
\label{eq:c}
c=2.997\,924\,58~10^5~{\rm km}\,{\rm s}^{-1}=
3.063\,915\,37~10^5~{\rm kpc}\,{\rm Gyr}^{-1}~~;
\end{equation}
on the other hand, the value of the
gravitation constant is deduced from
Eqs.\,(\ref{eq:kpc})-(\ref{eq:Gyr}), as:
\begin{equation}
\label{eq:G}
G=(6.672\,59\mp0.000\,85)~{\rm g}^{-1}\,{\rm cm}^3\,{\rm s}^{-2}=
(44927.5\mp5.7)~{\rm M}_{10}^{-1}\,{\rm kpc}^3\,{\rm Gyr}^{-2}~~.
\end{equation}

Using Eqs.\,(\ref{eq:kpc}) and (\ref{eq:M10}),
the unit density reads:
\begin{leftsubeqnarray}
\slabel{eq:rhoa}
&& 1\,{\rm M}_{10}\,{\rm kpc}^{-3}=6.770\,3~10^{-22}~{\rm g}\,
{\rm cm}^{-3}~~;  \\
\slabel{eq:rhob}
&& 1\,{\rm g}\,{\rm cm}^{-3}=1.477\,0~10^{21}~{\rm M}_{10}\,{\rm kpc}^{-3}~~;
\label{seq:rho}
\end{leftsubeqnarray}
and the value of the mean solar density is:
\begin{equation}
\label{eq:rhos}
\overline{\rho}_\odot=1.411~{\rm g}\,{\rm cm}^{-3}=
2.083~10^{21}~{\rm M}_{10}\,{\rm kpc}^{-3}~~;
\end{equation}
where $\overline{R}_\odot=6.96~10^{10}$ cm
has been assumed.

Using Eqs.\,(\ref{eq:kpc})-(\ref{seq:v}),
the unit energy and the unit specific
energy read:
\begin{leftsubeqnarray}
\slabel{eq:ea}
&& 1\,{\rm M}_{10}\,{\rm kpc}^2\,{\rm Gyr}^{-2}=
1.904\,3~10^{10}~{\rm g}\,{\rm cm}^2\,{\rm s}^{-2}~~;  \\
\slabel{eq:eb}
&& 1\,{\rm g}\,{\rm cm}^2\,{\rm s}^{-2}=5.251\,2~10^{-11}~
{\rm M}_{10}\,{\rm kpc}^2\,{\rm Gyr}^{-2}~~;
\label{seq:e}
\end{leftsubeqnarray}
\begin{leftsubeqnarray}
\slabel{eq:esa}
&& 1\,{\rm kpc}^2\,{\rm Gyr}^{-2}=
9.573\,877\,719\,424\,114~10^9~{\rm cm}^2\,{\rm s}^{-2}~~;  \\
\slabel{eq:esb}
&& 1\,{\rm cm}^2\,{\rm s}^{-2}=1.044\,508\,849\,294\,298~10^{-10}~
{\rm kpc}^2\,{\rm Gyr}^{-2}~~;
\label{seq:es}
\end{leftsubeqnarray}
similarly, the unit angular momentum
and specific angular momentum read:
\begin{leftsubeqnarray}
\slabel{eq:Ja}
&& 1\,{\rm M}_{10}\,{\rm kpc}^2\,{\rm Gyr}^{-1}=
6.005\,5~10^{69}~{\rm g}\,{\rm cm}^2\,{\rm s}^{-1}~~;  \\
\slabel{eq:Jb}
&& 1\,{\rm g}\,{\rm cm}^2\,{\rm s}^{-1}=1.665\,1~10^{-70}~
{\rm M}_{10}\,{\rm kpc}^2\,{\rm Gyr}^{-1}~~;
\label{seq:J}
\end{leftsubeqnarray}
\begin{leftsubeqnarray}
\slabel{eq:ja}
&& 1\,{\rm kpc}^2\,{\rm Gyr}^{-1}=
3.019\,218\,077\,027\,732~10^{26}~{\rm cm}^2\,{\rm s}^{-1}~~;  \\
\slabel{eq:jb}
&& 1\,{\rm cm}^2\,{\rm s}^{-1}=3.312\,115\,834\,257\,490~10^{-27}~
{\rm kpc}^2\,{\rm Gyr}^{-1}~~;
\label{seq:j}
\end{leftsubeqnarray}
accordingly, velocities, energies,
and angular momenta, may be translated
from [cm\,g\,s] to [kpc\,M$_{10}$\,Gyr]
system of measure and vice versa.

\section{Evolution of bound overdensities in CDM universes}
\label{a:g}

Overdensity evolution in CDM universes
is determined by Eqs.\,(\ref{eq:delRr}),
(\ref{eq:cir}), (\ref{seq:alt}),
(\ref{seq:alec}), and (\ref{seq:mov})-(\ref
{seq:gd}).   To perform calculations,
an explicit expression of the function,
$g(r)$, defined by Eq.\,(\ref{eq:gdb}),
is needed.   To help the reader, the
procedure used in earlier attempts
(Andriani and Caimmi, 1991, 1994)
shall be repeated here for the case
of interest i.e. bound perturbations
in an open universe.

To this aim, it is useful to define
the dimensionless variables (Andriani
and Caimmi, 1991):
\begin{leftsubeqnarray}
\slabel{eq:altpa}
&& \alpha^\prime=\frac r{r_i}~~;\qquad\tau^\prime=\frac t{(t_{ff})_i}~~; \\
\slabel{eq:altpb}
&& (t_{ff})_i=\left(\frac{\pi^2}8\frac{r_i^3}{G(M_m)_h}\right)^{1/2}=
\left(\frac{3\pi}{32}\frac1{G\rho_{hi}}\right)^{1/2}~~;\qquad\alpha_i^
\prime=1~~;
\label{seq:altp}
\end{leftsubeqnarray}
(the index, $h$, denotes zero density
excess) which are related to their
counterparts, expressed by Eqs.\,(\ref
{seq:alt}), via Eq.\,(\ref{seq:coi}) as:
\begin{equation}
\label{eq:aattp}
\alpha=\Delta_i\alpha^\prime~~;\qquad\tau=(1+\overline{\delta}_i)^{1/2}
\Delta_i^{3/2}\tau^\prime~~;
\end{equation}
with regard to an infinitely thin
spherical shell of initial radius,
$r_i$.

In addition, both the cosmic time,
$t$, and the free-fall time at the
beginning of evolution, $(t_{ff})_i$,
via Eqs.\,(\ref{seq:coi}) are independent
of $r_i$, which
implies null first derivatives:
\begin{equation}
\label{eq:dta0}
\frac{\partial\tau^\prime}{\partial r_i}=\frac{\partial\tau_i^\prime}
{\partial r_i}=0~~.
\end{equation}
Using Eqs.\,(\ref{seq:altp}), (\ref
{eq:aattp}), and the trigonometric
identity:
\begin{lefteqnarray*}
&& \frac12\arcsin(1-2x)+\frac\pi4=\arcsin\sqrt{1-x}~~;\qquad0\le x\le1~~;
\end{lefteqnarray*}
the following expression of
Eq.\,(\ref{eq:alw0}) holds:
\begin{lefteqnarray}
\label{eq:tap}
&& \tau^\prime-\tau_i^\prime=\mp\frac2\pi(1+\overline{\delta}_i)^{-1/2}
\Delta_i^{-3/2}\left[-\sqrt{\Delta_i\alpha^\prime-(\Delta_i\alpha^\prime)^2}
+\sqrt{\Delta_i-\Delta_i^2} \right. \nonumber \\
&& \phantom{\tau^\prime-\tau_i^\prime=}\left.+
\frac12\arccos(1-2\Delta_i\alpha^\prime)-\frac12
\arccos(1-2\Delta_i)\right]~~;
\end{lefteqnarray}
where the plus is related to expansion,
and the minus to contraction.

The combination of Eqs.\,(\ref{eq:gdb})
and (\ref{eq:altpa}) yields:
\begin{equation}
\label{eq:gp}
g(r)=\alpha^\prime\left[\frac{\partial(\alpha^\prime r_i)}{\partial r_i}
\right]^{-1}=\left(1+\frac{r_i}{\alpha^\prime}\frac{\partial\alpha^\prime}
{\partial r_i}\right)^{-1}~~;
\end{equation}
in terms of the dimensionless radius,
$\alpha^\prime$.

For spherical-symmetric overdensities,
Eq.\,(\ref{eq:delm}) reads:
\begin{equation}
\label{eq:demde}
\overline{\delta}=\frac3{r^3}\int_0^r\delta(r)r^2\diff r~~;
\end{equation}
and a derivation with respect to
$r$ produces:
\begin{equation}
\label{eq:ddemde}
\frac{\partial\overline{\delta}}{\partial r}=-\frac3r\overline{\delta}
\left(1-\frac\delta{\overline{\delta}}\right)~~.
\end{equation}

Keeping in mind that the density
parameter, $(\Omega_m)_i$, is
independent of $r_i$, a derivation
with respect to $r_i$ on both sides
of Eq.\,(\ref{eq:Rmaxb}) yields:
\begin{equation}
\label{eq:dDel}
\frac{\partial\Delta_i}{\partial r_i}=\frac{\partial\alpha_i}{\partial r_i}=
-\frac{-1+\Delta_i}{1+\overline{\delta}_i}\frac{\partial{\overline{\delta}_i}}
{\partial r_i}~~;
\end{equation}
owing to Eqs.\,(\ref{seq:Rmax}) and
(\ref{eq:alta}), where the former
has been generalized to the sphere bounded
by a generic isopycnic surface of
initial radius, $r_i$.

Using Eqs.\,(\ref{seq:altp})-(\ref{eq:dta0})
and (\ref{eq:gp})-(\ref{eq:dDel}),
a derivation with respect to $r_i$
on both sides of Eq.\,(\ref{eq:tap}),
after a lot of algebra, produces the
following expression of $g(r)$:
\begin{leftsubeqnarray}
\slabel{eq:gpia}
&& [g(r)]^{-1}=1+3\frac{1-\Delta_i}{\Delta_i}\frac{\overline{\delta}_i}
{1+\overline{\delta}_i}\left(1-\frac{\delta_i}{\overline{\delta}_i}
\right)\left\{1-\frac{\Delta_i^2}{\sqrt{\Delta_i-\Delta_i^2}}\frac
{\mp\sqrt{\Delta_i\alpha^\prime-(\Delta_i\alpha^\prime)^2}}{(\Delta_i
\alpha^\prime)^2}\times\right. \nonumber \\
&& 
\left.\times
\left[1+\frac{3\pi}4(1+\overline{\delta}_i)^{1/2}\left(\frac{1-\Delta_i}
{\Delta_i^3}\right)^{1/2}\left(1+\frac13\frac{\Delta_i}
{1-\Delta_i}\right)(\tau^\prime-\tau_i^\prime)\right]\right\}~~; \\
\slabel{eq:gpib}
&& [g(r_i)]^{-1}=1~~;
\label{seq:gpi}
\end{leftsubeqnarray}
for further details and exhaustive
investigation on open, flat, and
closed CDM universes hosting bound,
zero-energy, and unbound overdensities,
refer to Andriani and Caimmi (1991).

Turning again to the earlier
dimensionless coordinates,
$\alpha$ and $\tau$, the
combination of  Eqs.\,(\ref
{seq:alt}), (\ref{eq:aattp}),
and (\ref{seq:gpi}) yields
Eq.\,(\ref{eq:ges}).


\begin{thebibliography}{}
\bibitem{} Adams, F.C., Bloch, A.M., 2006. ApJ 653, 905.
\bibitem{}Ambartsumian, V., 1958. In {\it La Structure et l'Evolution de
          l'Universe}, ed. Institut International de Physique Solvay, p.\,267.
\bibitem{}Ambartsumian, V., 1965. In {\it The Structure and Evolution of
          Galaxies}, ed. Interscience Publishers, p.\,1.
\bibitem{} Amendola, L., 2000. PhRvD 62, 043511.
\bibitem{} Andriani, E., Caimmi, R., 1991. ApSS 182, 261.
\bibitem{} Andriani, E., Caimmi, R., 1994. A\&A 289, 1.
\bibitem{} Ascasibar, Y., Hoffman, Y., Gottl\"ober, S., 2007. MNRAS 376, 393.
\bibitem{} Babelberg, S., Shapiro, S.L., 2002. PhRvL 88, 101301.
\bibitem{} Baes, M., Buyle, P., Hau, G.K.T., Dejonghe, H., 2003. MNRAS 341,
           L44.
\bibitem{} Basilakos, S., 2003. ApJ 590, 636.
\bibitem{} Basilakos, S., Voglis, N., 2007. MNRAS 374, 269.
\bibitem{} Battye, R.A., Weller, J., 2003. PhRvD 68, 083506.
\bibitem{} Begelman, M.C., Rees, M.J., 1978. MNRAS 185, 847.
\bibitem{} Bett, P., et al., 2007. MNRAS 376, 215.
\bibitem{} Bromley, J.M., Somerville, R.S., Fabian, A.C., 2004. MNRAS 350,
           456.
\bibitem{} Brosche, P., Caimmi, R., Secco, L., 1983. A\&A 125, 338.
\bibitem{} Caimmi, R., 1989. A\&A 223, 29.
\bibitem{} Caimmi, R., 1990. A\&A 239, 7.
\bibitem{} Caimmi, R., 2006. Appl. Math. Comput. 174, 447.
\bibitem{} Caimmi, R., 2007. NewA 12, 327.
\bibitem{} Caimmi, R., Secco, L., Brosche, P., 1984. A\&A 139, 411.
\bibitem{} Caimmi, R., Secco, L., Andriani, E., 1990. ApSS 168, 131.
\bibitem{} Caimmi, R., Secco, L., 1992. ApJ 395, 119.
\bibitem{} Caimmi, R., Marmo, C., 2003. NewA 8, 119.
\bibitem{} Caimmi, R., Marmo, C., 2004. SerAJ 169, 11.
\bibitem{} Caimmi, R., Marmo, C., Valentinuzzi, T., 2005. SerAJ 170, 13.
\bibitem{} Caldwell, R.R., Dave, R., Steinhardt, P.J., 1998a. PhRvL 80, 1582.
\bibitem{} Caldwell, R.R., Dave, R., Steinhardt, P.J., 1998b. ApSS 261, 303.
\bibitem{} Carroll, S.M., Press, W.H.., Turner, E.L., 1992. ARA\&A 30, 499.
\bibitem{} Cavaliere, A., Morrison, P., Pacini, F., 1970. ApJ 162, L133.
\bibitem{} Cavaliere, A., Pacini, F., Setti, G., 1969. ApL 4, 103.
\bibitem{} Ciotti, L., Ostriker, J.P., 2007. ApJ 665, 1038.
\bibitem{} Colgate, S.A., 1967. ApJ 150, 163.
\bibitem{} David, L.P., Durisen, R.H., Cohn, H.N., 1987a. ApJ 313, 556.
\bibitem{} David, L.P., Durisen, R.H., Cohn, H.N., 1987b. ApJ 316, 505.
\bibitem{} Diemand, J., Moore, B., Stadel, J., 2004. MNRAS 353, 624.
\bibitem{} Di Matteo, T., Croft, R.A.C., Springel, V., Hernquist, L., 2003.
           ApJ 593, 56.
\bibitem{} Doran, M., Schmidt, J.-M., Wetterich, C., 2001. PhRvD 64, 123520.
\bibitem{} Duncan, M.J., Shapiro, S.L., 1983. ApJ 268, 565.
\bibitem{} Ferrarese, L., 2002. ApJ 578, 90.
\bibitem{} Ferrarese, L., Ford, H., 2005. Space Sci. Rev. 116, 523.
\bibitem{} Fukushige, T., Makino, J., 2001. ApJ 557, 533.
\bibitem{} Fukushige, T., Makino, J., 2003. ApJ 588, 674.
\bibitem{} Fukushige, T., Kawai, A., Makino, J., 2004. ApJ 606, 625.
\bibitem{} Granato, G.L., De Zotti, G., Silva, L., et al., 2004. ApJ 650, 42.
\bibitem{} Gunn, J.E., 1977. ApJ 218, 592.
\bibitem{} Gunn, J.E., 1987. {\it The Galaxy}, NATO ASI Ser.\,C207,
\bibitem{} Haehnelt, M.G., 2003. In Ho, L.C., ed, Carnegie Observatories
           Astrophys. Ser. Vol. I, {\it Coevolution of Black Holes and
           Galaxies}, Cambridge Univ. Press, Cambridge (astro-ph/0307378).
\bibitem{} Haehnelt, M.G., Rees, M.J., 1993. MNRAS 263, 168.
\bibitem{} Haiman, Z., Loeb, A., 2001. ApJ 552, 459.
\bibitem{} Heavens, A.F., Peacock, J.A., 1988. MNRAS 232, 339.
\bibitem{} Hernquist, L., 1990. ApJ 356, 359.
\bibitem{} Hiotelis, N., 2003. MNRAS 344, 149.
\bibitem{} Horellou, C., Berge, J., 2005. MNRAS 360, 1393.
\bibitem{} Iliev, I.T., Shapiro, P.R., 2001. MNRAS 325, 468. 
\bibitem{} Johnson, J.L., Bromm, V., 2007. MNRAS 374, 1557.
\bibitem{} Khlopov, M.Yu., Rubin, S.G., Sakharov, A.S., 2002. Grav. \&
               Cosmol. 8, Suppl., 57.
\bibitem{} Khlopov, M.Yu., Rubin, S.G., 2004. {\it Cosmological Pattern of
               Microphysics in the Inflationary Universe}, Kluver Academic  
               Publishers, Dordrecht.
\bibitem{} Khlopov, M.Yu., Rubin, S.G., Sakharov, A.S., 2005. Astropart.
               Phys. 23, 265.
\bibitem{} Koushiappas, S.M., Bullock, J., Dekel, A., 2004. MNRAS 354, 292.
\bibitem{} Landau, L., Lifchitz, E., 1966. {\it Theorie du champs}, Mir,
           Moscow.
\bibitem{} Lapi, A., Shankar, F., Mao, J., et al., 2006. ApJ 650, 42.
\bibitem{} Li, Y., Hernquist, L., Robertson, B., et al., 2007. ApJ 665, 187.
\bibitem{} Limber, D.N., 1959. ApJ 130, 414.
\bibitem{} Maccarone, T.J., Kundu, A., Zepf, S.E., et al., 2007.
           Nature 445, 183.
\bibitem{} Mack, K.J., Ostriker, J.P., Ricotti, M., 2007. ApJ 665, 1277.
\bibitem{} Madau, P., Rees, M.J., 2001. ApJ 551, L27.
\bibitem{} Maio, U., Dolag, K., Meneghetti, M., et al., 2006. MNRAS 373, 869.
\bibitem{} Manera, M., Mota, D.F., 2006. MNRAS 371, 1373.
\bibitem{} Maor, E., 1998. ApJ 494, L181.
\bibitem{} Maor, I., Lahav, O., 2005. JCAP 7, 3.
\bibitem{} Merloni, A., Nayakshin, S., Sunyaev, R.A., eds 2005. {\it Growing
         Black Holes: Accretion in a Cosmological Context}, Springer, Germany.
\bibitem{} Merritt, D., Navarro, F.J., Ludlow, A., et al., 2005. ApJ 624, L85.
\bibitem{} Merritt, D., 2006. Rep. Prog. Phys. 69, 2513.
\bibitem{} Miller, M.C., 2006. MNRAS 367, L32.
\bibitem{} Moore, B., Governato, F., Quinn, T., et al., 1998. ApJ 499, L5.  
\bibitem{} Moore, B., Quinn, T., Governato, F., et al., 1999. MNRAS 310,
           1147.
\bibitem{} Morrison, P., 1969. ApJ 157, L73.
\bibitem{} Mota, D.F., van de Bruck, C., 2004. A\&A 421, 71.
\bibitem{} Munyaneza, F., Viollier, R.D., 2002. ApJ 564, 274.
\bibitem{} Munyaneza, F., Biermann, P.L., 2006. A\&A 458, L9.
\bibitem{} Navarro, J.F., Frenk, C.S., White, S.D.M., 1995. MNRAS 275, 720.
\bibitem{} Navarro, J.F., Frenk, C.S., White, S.D.M., 1996. ApJ 462, 563. 
\bibitem{} Navarro, J.F., Hayashi, E., Power, C., et al., 2004. MNRAS 349,
           1039.
\bibitem{} Nomoto, K., Tominaga, N., Umeda, H., et al., 2005. Nucl. Phys. 32,
           51.
\bibitem{} Nunes, N.J., Mota, D.F., 2006. MNRAS 368, 751.
\bibitem{} Ohkubo, T., Umeda, H., Maeda, K., et al., 2006. ApJ 645, 1352.
\bibitem{} Peebles, P.J.E., 1980. {\it The Large-Scale Structure of
             the Universe}, Princeton Univ. Press, New Jersey.
\bibitem{} Pelupessy, F.I., Di Matteo, T., Ciardi, B., 2007. ApJ 665, 107.
\bibitem{} Percival, W.J., 2005. A\&A 443, 819.
\bibitem{} Portegies Zwart, S.F., Baumgardt, H., Hut, P., et al., 2004.
           Nature 428, 724.
\bibitem{} Quinlan, G.D., Shapiro, S.L., 1987. ApJ 321, 199.
\bibitem{} Quinlan, G.D., Shapiro, S.L., 1989. ApJ 343, 725.
\bibitem{} Reed, D., Governato, F., Verde, L., et al., 2005. MNRAS 357, 82.
\bibitem{} Rees, M.J., 1966. Nature 211, 468.
\bibitem{} Rees, M.J., 1984. ARA\&A 22, 471.
\bibitem{} Rubin, S.G., Sakharov, A.S., Khlopov, M. Yu., 2001. J. Exp. Theor.
               Phys. 92, 921.
\bibitem{} Ryden, B.S., Gunn, J.E., 1987. ApJ 318, 15.
\bibitem{} Sanders, R.H., 1970. ApJ 162, 791.
\bibitem{} Sersic, J.L., 1968. {\it Atlas de Galaxies Australes},
           Cordoba: Obs. Astron., Univ. Nac. Cordoba.
\bibitem{} Shankar, F., Mathur, S., 2007. ApJ 660, 1051.
\bibitem{} Shankar, F., Lapi, A., Salucci, P., et al., 2006. ApJ 643, 14.
\bibitem{} Silk, J., 2005. MNRAS 364, 1337.
\bibitem{} Silk, J., Rees, M.J., 1998. A\&A 331, L1.
\bibitem{} Silveira, V., Wega, I., 1994. PhRvD 64, 4890.
\bibitem{} Spaans, M., Silk, J., 2006. ApJ 652, 902.
\bibitem{} Spitzer, L.J., Saslaw, W.C., 1966. ApJ 143, 400.
\bibitem{} Spitzer, L.J., Stone, M.E., 1967. ApJ 147, 519.
\bibitem{} Terrel, J., 1964. Science 145, 918.
\bibitem{} Viollier, R.D., 1994. Prog. Part. Nucl. Phys. 32, 51.
\bibitem{} Volonteri, M., Haardt, F., Madau, P., 2003. ApJ 582, 559.
\bibitem{} Wang, L., Steinhardt, P.J., 1998. ApJ 508, 483.
\bibitem{} Wang, L., Caldwell, R.R., Ostriker, J.P., et al., 2000.
           ApJ 530, 17.
\bibitem{} Weinberg, N.N., Kamionkowski M., 2003. MNRAS 341, 251.
\bibitem{} Willott, C.J., Mclure, R.J., Jarvis, M.J., 2003. ApJ 587, L15.
\bibitem{} Zeldovich, J.B., Novikov, I.D., 1982. {\it Struttura
             ed Evoluzione dell'Universo}, Mir, Moscow.
\bibitem{} Zhao, H.S., 1996. MNRAS 278, 488.
\end{thebibliography}
\end{document}